\documentclass[aps,preprint,a4paper,prc,showpacs,showkeys,
nofootinbib]{revtex4-1}
\usepackage{graphicx}
\usepackage{bm} 
\usepackage{amssymb,amsmath,amsfonts}

\begin{document}

\title{NN scattering with $N\Delta$ coupling: Dibaryon
resonances without ``dibaryons''?}
\author{J. A. Niskanen}

\affiliation{Helsinki Institute of Physics, PO Box 64,
FIN-00014 University of Helsinki,  Finland
}
\email{jouni.niskanen@helsinki.fi}
\date{\today}

\begin{abstract}
It is known  that at their threshold intermediate
two-baryon $N\Delta$ states can produce resonance-like 
structures in some isospin one states, often interpreted 
as more exotic manifest six-quark states. 
This paper applies the coupled-channel method to study vital
details of the $N\Delta$ effect in isospin one $NN$ scattering
with interactions constrained to exclude the influence of
such extraneous hypothetical particles in two respects.
Firstly, the calculated phase parameters are fitted only below 
pion inelasticity, i.e. far below the region of ``dibaryons''. 
Secondly, the strength of the $NN \rightarrow N\Delta$ 
transition is
constrained to agree with the pion production reaction
$NN\rightarrow d\pi$, fairly independent of the details of
the pure ``diagonal'' $NN$ potential and its effect practically 
as strong as the $NN$ itself. The strongly angular momentum
dependent ingredients and results may be 
considered as a necessary background for searches of any
additional dibaryon effects.
Optimal conditions for resonant $N\Delta$
appearances involve a decrease of the orbital
angular momentum in the transition $NN\rightarrow N\Delta$,
as is the case in the $^1D_2$ and $^3F_3$ waves.
Detailed complex phase shift results are further presented also 
for isospin one peripheral partial waves up to $J=6$, where,
contrary to common prejudice, $N\Delta$ 
excitation is still found to be comparable to or even larger 
than one pion exchange.
\end{abstract}

\maketitle

\section{Introduction}  \label{intro}
Shortly after the appearance of the quark model in
1964 \cite{gellmann} it was suggested \cite{dyson,jaffe} the 
possible existence also of states with six quarks (somewhat 
like six-quark bags). Although this idea concerned only
color singlet involving all three quark flavors
(known at the time), as a follow-up it was further
suggested that this kind of objects would play an important 
role as resonant $s$-channel intermediate states in the 
interaction 
dynamics of two nucleons (for a balanced review see e.g. Ref.
\cite{dibaryn}). Searches have since continued for
indications in
experimental data (for an early review in the $NN$ sector 
see e.g. Ref. \cite{yokosawa}). An interest in such activity
has been sustained over decades and got new wind from 
observations at the WASA@COSY detector of Forschungszentrum
J\"ulich in two-pion production reactions, where a resonant
structure with $I(J^P) = 0(3^+)$, called $d'(2380)$, 
was seen at 2380 MeV with the width of 70 MeV \cite{adlar106}.

However, there is another well established possibility for strong 
resonance-like energy dependencies, namely two-pion exchange
with the excitation of an intermediate state with a 
 $\Delta(1232)$ resonance. It was
realized early in pion physics that pion-nucleon interaction,
both elastic scattering and reactions, in the region of a
few hundred MeV is dominated by this spin-$\frac{3}{2}$ and
isospin-$\frac{3}{2}$ particle \cite{watson}. Pion exchange can 
naturally induce such an excitation in isospin one $NN$ 
scattering \cite{sugawara}, transforming one of the nucleons 
or both into a $\Delta$. Then its consequent decay may end up
in pion production reaction or (if reabsorbed by the other 
nucleon) generate strong attractive two-pion exchange, which is 
not reducible to two-nucleon intermediate states appearing 
in iteration of a normal $NN$ potential \cite{green,arenhovel}.
This mechanism acts in the same energy region as several of
the conjectured
dibaryons and may compete with their effects or be interpreted
as those. In the 1970's and 1980's the attractive effect of 
$\Delta$ excitation was
included for example in the so called Bonn boson exchange potentials
\cite{machleidt,elster}, well fitted to phase shift analyses
in the elastic energy region. The aim of the present paper is to 
elaborate the coupled channels model more in the inelastic region.

The ascent of 
inelasticity above pion threshold was studied long ago
by $N\Delta$ coupled channels in $NN$ scattering
e.g. in Refs. \cite{gns} and \cite{gs}. The positions of 
possible associated
resonant poles were studied by VerWest \cite{verwest} 
and Kloet and Tjon \cite{kloet} in some detail using
exactly solvable separable
models, which showed the mathematical reality of such 
threshold effects in $N\Delta$ states. 
Also Betz and Lee had studied pion-nucleon and pion-deuteron 
dynamics, albeit also with separable interactions \cite{betz}.
Van Faassen and Kloet went further to devise realistic
interactions with meson exchanges in Ref. \cite{faassen} and
produced qualitatively successful phases and inelasticities 
also in the dibaryon regime at and above the $N\Delta$ threshold. Likewise, the Bonn potential was successfully extended
to include inelasticity for applicability at the dibaryon 
energies \cite{elsternew}. 

However, these realistic interactions \cite{faassen,elsternew}
are still mere $NN$ potentials in the sense that they start
with $NN$ and end with $NN$, producing no $N\Delta$ wave
function necessary to calculate reactions and other probes of
the intermediate state.
Neither do the others above. This is the crucial difference 
to the present paper. Here, in scattering calculations the 
transition $NN \rightarrow N\Delta$ will be given also a strict
external constraint arising from the total cross section
of the reaction $NN \rightarrow d\pi$, which requires and is
sensitive to such a wave function and, thus, can fix the 
strength of that transition.

The next section outlines the basic formalism for generating 
the $N\Delta$ configurations in the isovector $NN$ scattering
and discusses their main properties. First the coupled
Schr\"odinger equation is given in Subsection \ref{ccm}
and then Subsec. \ref{transition} introduces its most
important and essential ingredient, the transition potential.
Rise of inelasticity in terms of the effective width of
the $N\Delta$ channel is discussed in Subsection \ref{inelrise}
and the potential present in the incident $NN$ channel 
in \ref{nnsector}. Finally some additional comments on the
general formalism are given in Subsec. \ref{additional}.
The numerical results for the $NN$ phase shifts are 
presented and discussed partial wave by partial wave in 
Sec. \ref{results} before summarizing in Sec. \ref{conclu}.

\section{$N\Delta$ states}   \label{single}
 \subsection{Coupled channels system}   \label{ccm}

In this work (as in its predecessors since Ref. \cite{ppdpi}
and the more mature Ref. \cite{improv}) the
$N\Delta$ components $w_i(r)$ are generated from the incident 
$NN$ wave $u(r)$ by a coupled system of Schr\"odinger 
equations involving as the new element the
transition potential $NN \leftrightarrow \Delta N$ due mainly
to pion exchange
\begin{eqnarray}
  & \displaystyle
  \left[ -\frac{{\rm d}^2}{{\rm d}r^2} + \frac{L(L+1)}{r^2}
 - k^2 + \frac{M}{\hbar^2} V_{\rm NN} \right] u(r) \nonumber \\
 & \displaystyle     \label{nnchan}
  = - \frac{M}{\hbar^2} \sum_i V_{\rm tr}(i)\, w_i(r) \, , \\
    & \displaystyle
  \left[ -\frac{{\rm d}^2}{{\rm d}r^2} + \frac{L'_i(L'_i+1)}{r^2}
 - k^2 + \frac{M'}{\hbar^2} (\Delta M+V_{\rm N\Delta}(i)) \right] 
 w_i(r) \nonumber \\
 & \displaystyle    \label{ndchan}
  = - \frac{M'}{\hbar^2}  V_{\rm tr}(i)\, u(r) \, .
\end{eqnarray}
Here $k$ is the centre of mass momentum, $L$ ($L'_i$) the
$NN$ ($N\Delta$) angular momentum, $M$ and $M'$ twice the  
respective reduced masses. The mass difference $\Delta M$
between the $\Delta$ and nucleon defines the threshold of
real $\Delta$'s. For the $\Delta$ mass the real part of the
position of the pole is used giving $\Delta M = 274$ MeV.

There is another lower 
important threshold already halfway from 
real $\Delta$'s: namely pion production threshold. This 
inelasticity can and will be treated by an imaginary term, 
half-width in the $\Delta$ mass eventually showing up in 
the imaginary parts of $NN$ phase shifts.
However, for the reasons pointed out later in Subsection  
\ref{inel} this cannot be the free $\Delta$ width.
It will be discussed closer in Subsec. \ref{inelrise}.

Eqs. (\ref{nnchan}) and (\ref{ndchan})
give an impression of quite a lot of freedom in the system,
involving altogether three interactions and the width, 
probably too much to fit by $NN$
scattering alone and to make  definitive conclusions about
dibaryons. However, these are somewhat interrelated and it
is possible also to find some external constraints as seen
in the next two Subsections.

First, an interesting possibility would be attraction 
between the nucleon and $\Delta$, conceivably strong enough 
for a bound state \cite{arbound}. 
However, the diagonal $V_{N\Delta}$ is somewhat far 
removed from $NN$, as a higher order effect than
 $V_{NN}$ or pure excitation of $\Delta$ by
$V_{\rm tr}$, and it is difficult to reach unambiguously 
by experiments. Diagonal one pion exchange 
$N\Delta\rightarrow N\Delta$ may be rather weak due to the small
pion coupling to $\Delta$ 
$f_{\pi \Delta\Delta} = f_{\pi NN}/5$ from the quark 
model \cite{brown}. In pion production and photoreactions 
its effect in two models has been modest or small 
 \cite{ppdpi,wilhelm}. Because the aim of this work 
is not to try to impose ``dibaryons'' (i.e. more exotic than
the present coupling to $N\Delta$ excitation)
but only to study minimally how the $NN$
phases behave with minimal assumptions, this interaction is
neglected and will not be discussed any more. Therefore, 
the possibly predicted resonant structures 
could be regarded as threshold cusps.

\subsection{$NN\rightarrow N\Delta$ coupling potential}
   \label{transition}
The familiar one pion exchange (OPE) potential 
between two nucleons can be naturally and straightforwardly 
generalized by the transition spin formalism featured e.g.
in Refs. \cite{sugawara,brown} into a transition 
potential of the form
\begin{equation}
V_{\rm tr} =  \frac{\mu}{3} \frac{f f^\ast}{4\pi}
   {\bm T}_1 \cdot {\bm \tau}_2 \, [S_{12}\, V_{\rm T}(r) +
   {\bm S}_1\cdot{\bm \sigma}_2\, V_{\rm SS}(r) ]
      + (1 \leftrightarrow 2)  \; .     \label{transpot}
\end{equation}
Here $f^2 /4\pi = 0.076$ and $f^{\ast 2}/4\pi = 0.35$
from the $\Delta$ width \cite{sugawara} are the pion
(pseudovector) coupling constants to the nucleon and
$N \leftrightarrow \Delta$ vertices, and $\mu$ is the pion mass. 
$ \bm S$ and ${\bm S}^\dag$ are the transition spin
operators \cite{brown} (analogously $\bf T$ and ${\bm T}^\dag$
for the isospin) defined for example
 by their reduced matrix elements
$<\Delta || {\bm S} || N > = 2$ in the convention of
Refs. \cite{edmonds,deshalit,messiah}. Here the standard 
tensor operator
\begin{equation}
S_{12} =  3\, {\bm \sigma}_1\cdot \hat{\bm r} 
            {\bm \sigma}_2 \cdot \hat{\bm r}
           -{\bm \sigma}_1 \cdot {\bm \sigma}_2
\end{equation}
has been generalized to
 \begin{equation}
S_{12} =   3\, {\bm S}_1\cdot \hat{\bm r} 
            {\bm \sigma}_2 \cdot \hat{\bm r}
           -{\bm S}_1 \cdot {\bm \sigma}_2 +(1\leftrightarrow 2)
\end{equation}
in the case of the $NN \rightarrow \Delta N$ transition. The radial
functions are  
\begin{eqnarray}
& V_{\rm SS}(r)   = \exp(-\mu r) /(\mu r)  \nonumber \\
& V_{\rm T}(r)  = [1 + 3/(\mu r) +3/(\mu r)^2 ]\, V_{\rm SS}(r)\, ,
 \label{range}
\end{eqnarray}
also familiar from pion exchange between two nucleons. The 
$NN$ equation is then coupled by $V_{\rm tr}$ to the $N\Delta$ equations including (in addition to their possible mutual 
interactions not discussed here) the $\Delta - N$ mass 
difference $\Delta M = 274$ MeV as given in the previous
Subsection and (above the inelastic threshold) the effect 
of the width $-i\Gamma_i /2$ in the $N\Delta$ channel $i$.

This procedure yields an admixture of $N\Delta$ 
configurations in principle on the same footing as the $NN$ 
(apart from the mass difference and asymptotic attenuation) 
into the wave function, which must necessarily be there due to 
the nucleon and and pion coupling to the $\Delta$. 
The method was applied 
relatively successfully to calculate both total and differential
cross sections and analyzing powers \cite{ppdpi} and other spin
observables \cite{spinobs,deutpol} for the reaction
$pp \rightarrow d\pi^+$. Significant improvements and 
complements have been included in Ref. \cite{improv}.
It may be useful to have in mind that 
the background of this paper lies in this reaction, and some
argumentation to follow largely arises and and is adopted from
that work and its later developments.

At this stage it may be worth commenting on the radial
dependencies (\ref{range}) that, 
following the suggestion by Durso {\it et al.} \cite{durso},
originally two time orderings in OPE
taking into account the different $\Delta$ mass was used for
the range in Ref. \cite{ppdpi}. However, inclusion of all 
time orderings gives back the ``normal'' static OPE 
range $1/\mu$ in Eq. (\ref{range}) used later since 
Ref. \cite{improv}. 

The transition potential presently also 
includes $\rho$ exchange and dipole form factors with cut-off 
masses $\Lambda_\pi = 1000$ MeV and $\Lambda_\rho = 1050$ MeV. 
Interfering destructively with pion exchange, $\rho$ exchange 
acts in the dominant tensor term $V_{\rm T}(r)$ similarly 
to a long range cut-off. The spin-spin part $V_{SS}(r)$ 
of the transition potential 
(\ref{transpot}) is a much weaker effect. Actually
for practical purposes, it would be nearly enough 
to get the strength of the dominant tensor part from the top 
of the peak of the 
$pp \rightarrow d\pi^+$ total cross section \cite{hoftiezer}
as the only adjusted parameter relevant for $N\Delta$.

Of course, the whole transition potential (\ref{transpot})
is used with the above form factors, well in accord with 
the nucleon size and mesonic corrections to the free 
$\Delta$ width \cite{pindcpl}. Supplemented by the
effective $N\Delta$ state width calculated in the next
Subsection it reproduces the $NN\rightarrow d\pi^+$ cross 
section excellently from the threshold to 800 MeV and 
a bit beyond \cite{improv}. This choice of the form factors
is the only fine-tuning of the model done in the 
``dibaryon'' region (i.e. above inelasticity threshold) and 
it is not sensitive or closely related to $NN$ scattering 
itself. Therefore, it is reasonable to adopt the 
transition potential defined above, acceptable for 
$pp\rightarrow d\pi^+$, also to $NN$ scattering. 
In this success another essential element is also the
treatment of the $N\Delta$ channel width discussed in the 
next Subsection. This is not the free $\Delta$ width but
it is not a free parameter either.

One may note that, although, due to the two-body 
phase space, this cross section is much smaller than the 
total inelasticity of $NN$ scattering, 
its data are very good and decisive \cite{hoftiezer}. 
Actually, its highest reported point 
$\sigma_{\rm prod} = 3.148$ mb at 575.2 MeV (laboratory energy)
is probably the most sensitive probe for the $N\Delta$ 
transition potential strength. (An interpolation of 
the data would give 3.157 mb at 577 MeV.)
In this way the uncertainty of the $N\Delta$ vs. $NN$ effect 
be reduced from even some tens of percent to few percent,
as might be inferred from Figs. 2 and 3 of Ref. \cite{wilhelm}.
The present transition potential defined above 
yields a flat maximum of about 3.285 mb between 583 and 
588 MeV, very close. Then the elastic $NN$ phase shift is 
fitted by $V_{NN}$ only {\em below} inelasticity.
No overall phase shift fit extending above 
pion threshold is attempted, so that the structures in the
``dibaryon'' region are due to the model, not from fits.

Heavy mesons like the $\rho$ may be frowned in more recent
chiral effective field theories (EFT), composed as systematic
pion exchange perturbation series including $\pi N$ and $\pi\pi$
contact terms with few low-energy parameters. (See e.g.
\cite{eft} for recent progress and extensive bibliography.)
However, so far this approach has been limited to lower $NN$
momenta below inelasticity. Therefore, a more phenomenological
approach consisting of fewer plausible iterative diagrams, 
but extending higher in energy, may have value and interest.
Concerning the role of $\rho$ exchange, it might be of some 
interest in future work to compare spin observables of 
$pp \rightarrow d\pi^+$ including and omitting $\rho$ 
exchange to test the justification of its inclusion, 
since, contrary to the tensor part, its spin-spin  term 
(albeit much weaker) adds constructively with 
pion exchange, which should give differing spin dependence.

\subsection{Rise of inelasticity}    \label{inelrise}

The imaginary term from the $\Delta$ width gives naturally 
also a prediction for $NN$ inelasticity above pion threshold. 
There are some subtleties required in its inclusion.
In Ref. \cite{ppdpi} a simple kinematic adjustment is applied on
the free $\Delta$ width. (Also in \cite{faassen,elsternew} 
different state-independent prescriptions were used.) 

However, it is important to realize that, because different 
dynamics in different $N\Delta$ channels yield different 
wave functions (and therefore different momentum 
distributions), also widths are affected. 
In particular, the  baryon kinetic energy should not be 
available for the $\Delta$ decay but should be subtracted
from the total energy by kinematics.
The effective $N\Delta$ width
taking into account the relative kinetic energy of the nucleon
and $\Delta$ is in the momentum representation \cite{improv,width}
\begin{equation}
\Gamma_3 (i) = \frac{2}{\pi}\,
\frac{\int_0^{p_{\rm max}} |\Psi_{N\Delta}(p)|^2\,
\Gamma(q)\, p^2\, dp}
{\int_0^\infty |\Psi_{N\Delta}(r)|^2\, r^2\, dr} \, .
\label{gamma3}
\end{equation}
Here $\Psi_{N\Delta}(p)$ is the Fourier transform 
of the appropriate $N\Delta$ channel wave function 
$\Psi_{N\Delta}(r) = w_i(r)$
and $\Gamma(q)$ the free $\Delta \rightarrow N\pi$ 
width \cite{bransden}
\begin{equation}
\Gamma(q) = \frac{142\, (0.81\, q/\mu)^3}
{1+(0.81\, q/\mu)^2} \; {\rm MeV}
\label{free}
\end{equation}
with $q$ as the relative $N\pi$ momentum compatible with 
the total energy and baryon (c.m.) momentum $p$. 
The subscript 3 refers to three-body decay 
$N\Delta\rightarrow NN\pi$, the main inelasticity.
In the calculations of the width and the associated
scattering also the calculated 
two-body inelasticity $d\pi$ is taken into account as 
described in Refs.  \cite{improv,width}.

Because in Eq. (\ref{gamma3})
the wave function influences the channel interaction, which
in turn yields the wave function, the procedure requires 
self-consistent iterative calculations for this 
inelasticity generating agent. The width decreases quite
substantially from its free value, more for higher 
orbital angular momenta of the $N\Delta$ system \cite{width}.
The main effect to the cross section is increase, because the
the absorption acts much like effective repulsion.
The result is a significant 
improvement of the calculated observables \cite{improv}
and nearly perfect agreement with the total cross section
of $pp\rightarrow d\pi^+$. It should be noted that though
the success is based on both the transition potential and
width, the latter is not an arbitrary or fitted quantity but 
calculated from the well known $\Delta$ free width
(\ref{free}) and the only free fitted parameters of
significance to $pp\rightarrow d\pi^+$ are the form factors,
basically the height of the peak. 

Further, it may also be remarked that the width, a uniform 
imaginary potential
extending spatially to infinity, causes the wave function 
to fall off making the $N\Delta$ channel asymptotically closed
also above threshold as well as below, not unlike a bound state. 
A rotational series of energy states is not far fetched as will
be seen later in the next Subsection.

\subsection{Diagonal $NN$ potential}   \label{nnsector}

Apart from the explicit generation and use of the $N\Delta$ 
wave function with coupled channels, as iterated pion 
exchange $V_{\rm tr}(r)$ produces a strong 
attraction at least below the formal $N\Delta$ threshold, 
which must be accounted for in dealing with  the overall 
effective two-nucleon force $V_{NN}(r)$. 
Namely, phenomenological data-fitted $NN$ potentials, 
such as e.g. the Reid potential \cite{reid}, already 
by definition include this attraction provided by Nature 
itself and, therefore,
they cannot be $V_{NN}(r)$ in Eq. \ref{ccm}. A theorist
working with the equation must then subtract this attraction 
to avoid doubly counting its influence.

In Ref. \cite{ppdpi} this was done by assuming a closure
approximation for the iterated transition and using an
additive repulsive term
\begin{equation}
V_{\rm clos}(r) = [V_{\rm tr}(r)]^2 / \Delta E
\label{closure}
\end{equation}
with the energy denominator $\Delta E$ suitably adjusted 
to reproduce the phase shifts at each energy. It should 
be stressed that 
at that time the purpose was not to calculate or publish 
$NN$ scattering and its phases but only to generate dependable 
$N\Delta$ wave functions for computing the pion 
production matrix elements necessary in 
the reaction $pp \rightarrow d\pi^+$. As discussed before,
the knowledge of the $N\Delta$ component from this reaction
can be used for fixing the transition potential for coupled 
channels independently of pure $NN$ scattering as presented
above. It may be noted that, because the $NN$ potential has
to be adjusted anyway (and $pp\rightarrow d\pi^+$ is rather
insensitive to that), it does not matter much which 
phenomenological potential is used as a starting point.

In the present context of dibaryons it is also of interest to 
note that, to produce phase equivalent
interactions with and without the $N\Delta$ admixture
by this procedure, $\Delta E$ of Eq. (\ref{closure})
predicted a rotational series \cite{dibarmass}
$\Delta E \approx {\rm const} + 40 L_{N\Delta}(L_{N\Delta}+1)$
MeV for the effective intermediate state (isospin one ``dibaryon'') masses, well in 
accord with the contemporary experimental situation \cite{yokosawa}. Effectively this result could also be 
interpreted so that the $N\Delta$ state wave 
function acted as if 
it were approximately concentrated at the distance of about 
1 fm. This is also well commensurate with the OPE range.

Later the double-counting modification has been performed by 
inclusion of energy-independent short and
intermediate range Yukawa terms \cite{width,csb} into the Reid 
potential \cite{reid} and its extension to higher partial waves 
with $J=3$ \cite{day}. A short digression on these changes is
given in the end of this Subsection. From Eq. \ref{closure} 
and second order 
perturbation arguments it should be clear that the correction should be proportional to some average of the square of
the transition
strength and from experience that it is strong. And as given, 
the correction is energy independent.

It is worth reiterating that, as different as the above described methods to deal with double counting may appear, their 
principal application to the reaction $pp\rightarrow d\pi^+$
is rather independent of these details. In particular, its
total cross section is practically independent of the $NN$
phase shifts, whereas it is very sensitive to the $N\Delta$
component and thus to the $NN\rightarrow N\Delta$ 
transition strength.
 
An essential issue of this paper is the following: to obtain
from a seemingly energy-independent potential strong energy 
dependencies, often attributed to and parametrized and fitted 
with explicitly energy-dependent ``dibaryons''. As discussed
in Introduction, arguments to this end have been given long ago 
e.g. in Refs. \cite{verwest,kloet,betz,faassen,elsternew}. 
However, it is hard to say from these works (fixed by $NN$
scattering) which transition strength $V_{\rm tr}$ is 
correct, since the effect of its changes
can be countered in the elastic $NN$ sector.

Because of the interplay between the (complex) 
$N\Delta$ related attraction 
and the ``diagonal'' pure $NN$ potential, from the above
adjustments it may be seen that $NN$ scattering alone
is quite soft for an unambiguous sharing of the contributions 
from the $N\Delta$ and $NN$ parts. Namely, for example,
missing attraction can be provided by either increasing the 
strength of $V_{\rm tr}$ or inserting it directly in $V_{NN}$
and vice versa for repulsion.
Furthermore, increasing the width does not 
necessarily increase inelasticity linearly, since it also 
shrinks the $N\Delta$ component. Anyway, a weaker transition 
potential in turn has directly a side effect  e.g. as
lack of inelasticity in Ref. \cite{elsternew}, which uses
a weaker $\pi N\Delta$ coupling 
${f^\ast}^2/4\pi = (72/25)f^2/4\pi \approx 0.23$  from the
quark model.\footnote{This coupling would also underestimate 
$pp\rightarrow d\pi^+$
and photoabsorption on deuteron by a similar 
factor as seen in Figs. 2 and 3 of Ref. \cite{wilhelm}.}
In fact, the difference between the present $N\Delta\pi$
coupling ${f^\ast}^2/4\pi = 0.35$ can be accounted for by
vertex corrections \cite{pindcpl}.

Further, another complication in these analyses 
with complex potentials is the fact 
that the absorptive imaginary interaction term itself acts 
effectively as repulsion \cite{antipscatt}. Therefore, 
it is a great boon and benefit to have an additional
constraint for the transition strength rather independent of
$NN$ scattering as described above.

\subsubsection{Modification of $V_{NN}$}
For definiteness here the repulsive  modification
(in MeV) described above
is given to lower partial waves. These were fitted for
the Reid potential \cite{reid} and its extension to $^3F_3$
\cite{day}, but they should be reasonable at least in pion 
production for any modern phenomenological potential.
The radial variable is $x = \mu r = 0.7r\times\,$fm$^{-1}$.
The potential changes have been adjusted in the elastic
region to the data of Ref. \cite{arndt}. They have been very
stable below 1000 MeV, can be considered sufficiently precise
for the present purpose as will be seen in comparisons later.
\begin{equation}
\Delta V_{\rm C}(^1S_0) = 770 e^{-4x}/x
\end{equation}
\begin{equation}
\Delta V_{\rm C}(^3P_0) = 200 e^{-4x}/x
\end{equation}
\begin{equation}
\Delta V_{\rm C}(^3P_1) = -150 e^{-3x}/x + 18000 e^{-7x}/x
\end{equation}
\begin{eqnarray} 
 & \displaystyle
\Delta V_{\rm C}(^3P_2-^3F_2) = 180 e^{-3x}/x + 1200 e^{-6x}/x
 \nonumber  \\
 & \displaystyle \Delta V_{\rm T}(^3P_2-^3F_2) =  800 e^{-6x}/x 
 \nonumber  \\
 & \displaystyle
\Delta V_{\rm LS}(^3P_2-^3F_2) = 10 e^{-3x}/x - 400 e^{-6x}/x
\end{eqnarray}
\begin{equation}
\Delta V_{\rm C}(^1D_2) = 230 e^{-3x}/x + 8000 e^{-7x}/x
\end{equation}
\begin{equation}
\Delta V_{\rm C}(^3F_3) = 2700 e^{-5x}/x
\end{equation}
For $^1G_4$ wave initially the Reid potential for the $^1D_2$
state was used modified by
\begin{equation}
\Delta V_{\rm C}(^1D_2) = 300 e^{-3x}/x + 8000 e^{-7x}/x \; .
\end{equation}

\subsection{Additional comments}   \label{additional}
 
Dibaryon fits have often only considered a single
reaction (e.g. $NN$ scattering, the topic here) or a single
observable (typically a total cross section) 
without a careful study of $N\Delta$ and its ramifications. 
For example Kamo and Watari \cite{kamo} fitted
the total cross section of $pp \rightarrow d\pi^+$ with three
dibaryons (i.e. six parameters), assuming a constant background 
from the $\Delta$. However, the $N\Delta$ threshold effect does
not behave like a constant background and, in fact, the coupled 
$N\Delta$ model \cite{ppdpi}
reproduced both the total cross section and also differential
and spin observables quite well, even better than 
Ref. \cite{kamo}, particularly after some improvements 
in the treatment of pion $s$-wave rescattering and the $N\Delta$ 
width \cite{improv,width}. (To my knowledge dibaryon fits have 
not often been applied to differential observables in pion
production, though Ref. \cite{kamo} does give predictions or fits
with also these.)

It might appear that, in second order iteration of pion 
exchange in $NN$ scattering (and first order in pion production),
the effect of the intermediate $\Delta$ could be obtained simply
by inserting into the perturbation theory energy denominator 
the $\Delta\! -\! N$ 
mass difference $\Delta M$ and width
giving $E-\Delta M + i\Gamma/2$. Although this looks like a 
trivial way to get a resonant structure, however, it is open to 
some questions. It was pointed out above that already the 
extremely phenomenological numerical calculation \cite{dibarmass}
for phase equivalent potentials 
indicates some more subtle structure in the energy denominator,
e.g. from the centrifugal
effects discussed in more detail in Ref. \cite{centrif}. 
This is due to the fact that in asymptotically closed $N\Delta$
channels the expectation value of the centrifugal potential
is actually well defined and quantized as argued 
above in Subsec. \ref{inelrise}. Furthermore, 
the orbital angular momentum of the excitation can be 
$L_{NN}\pm 2$ in addition to $L_{NN}$, and the 
$N\Delta$ centrifugal barrier obviously favoring $L_{NN}-2$
and strongly suppressing $L_{NN}+2$.
Even the definition of the width in the two-baryon system needs 
special attention and becomes strongly state dependent. 
As seen, quite obviously it cannot be the free
width \cite{improv,width}, but consideration of the different 
parts of the $N\Delta$ kinetic energy  expectation values
should be addressed apart from the internal $\Delta$ particle
excitation as in Eq. (\ref{gamma3}).

Moreover, there is a perhaps deeper problem
of the applicability of perturbation theory in this context:
Perturbation theory needs the matrix element of the perturbation between the initial and intermediate (or final) state to
generate its effect. But what is the ``unperturbed'' wave function 
of the $N\Delta$, which has yet to be computed to
get this matrix element? This is a problem in old perturbation
calculations such as Ref. \cite{brack}, which can reproduce the
total cross section of $pp \rightarrow d\pi^+$ well by 
straightforward substitution of the above energy denominator,
but not the differential one or spin observables \cite{chai,maxwell1,maxwell2}.

The question of the applicability of perturbation theory may 
not be as relevant in $NN$ scattering for the fits with 
hypothesized dibaryons. The resonant structure as 
substituted by hand is then trivial and automatic. 
The positions and widths are free to choose in this
procedure. With the $N\Delta$ coupled channels
there is no such luxury for
these. The position and magnitude come directly from the 
coupled Schr\"odinger equation as a threshold cusp
(with the $\Delta M$ and centrifugal forces as necessary
and unavoidable 
parts of the equation), and the width is calculated 
self-consistently for each channel along the lines of 
Ref. \cite{width}.

\begin{figure}[tb]
\includegraphics[width=\columnwidth]{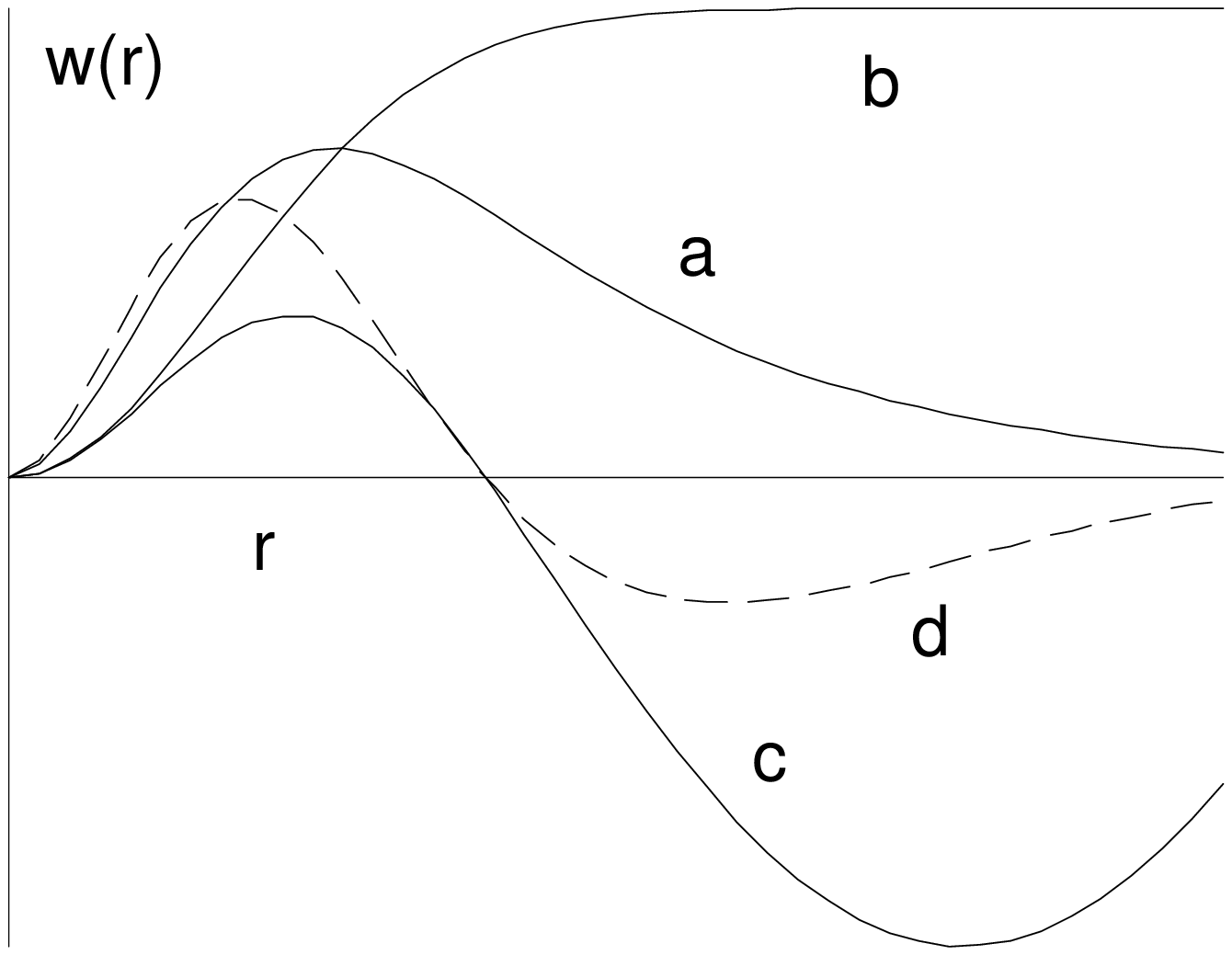}
\caption{Schematic representation of the $N\Delta$ wave
function around the channel threshold. Curves a, b and c
below, at and above threshold without width. Curve d depicts
the attenuation due to the width in the oscillatory case.
\label{demo}
}
\end{figure}

Although in the coordinate space equations the resonant  
structure is not manifest, in the equivalent momentum (or energy) 
representation, of course, it must be and, after calculation,
becomes visible for observables also in the 
coordinate representation. 
This is schematically demonstrated below and above the 
$N\Delta$ threshold in Fig. \ref{demo}. First, without
width curve a below threshold decays exponentially with $r$.
At exactly threshold it is constant outside the potential 
range (curve b), whereas above oscillatory behaviour arises 
(curve c). It is quite likely that in amplitudes the functions
a and c would yield smaller overlap integrals than b. 
Depending on final states and interactions this may not
always be true, but mostly this is the case for cusp maxima.
The dashed curve d shows the qualitative moderating effect 
of adding the width to curve c (only the possible real 
part shown). So, from this on we'll 
stick to this representation, which is also the most
transparent framework to incorporate the necessary 
centrifugal barriers.

\section{Results}   \label{results}
\subsection{$^1D_2$ wave}

\begin{figure}[tb]
\includegraphics[width=\columnwidth]{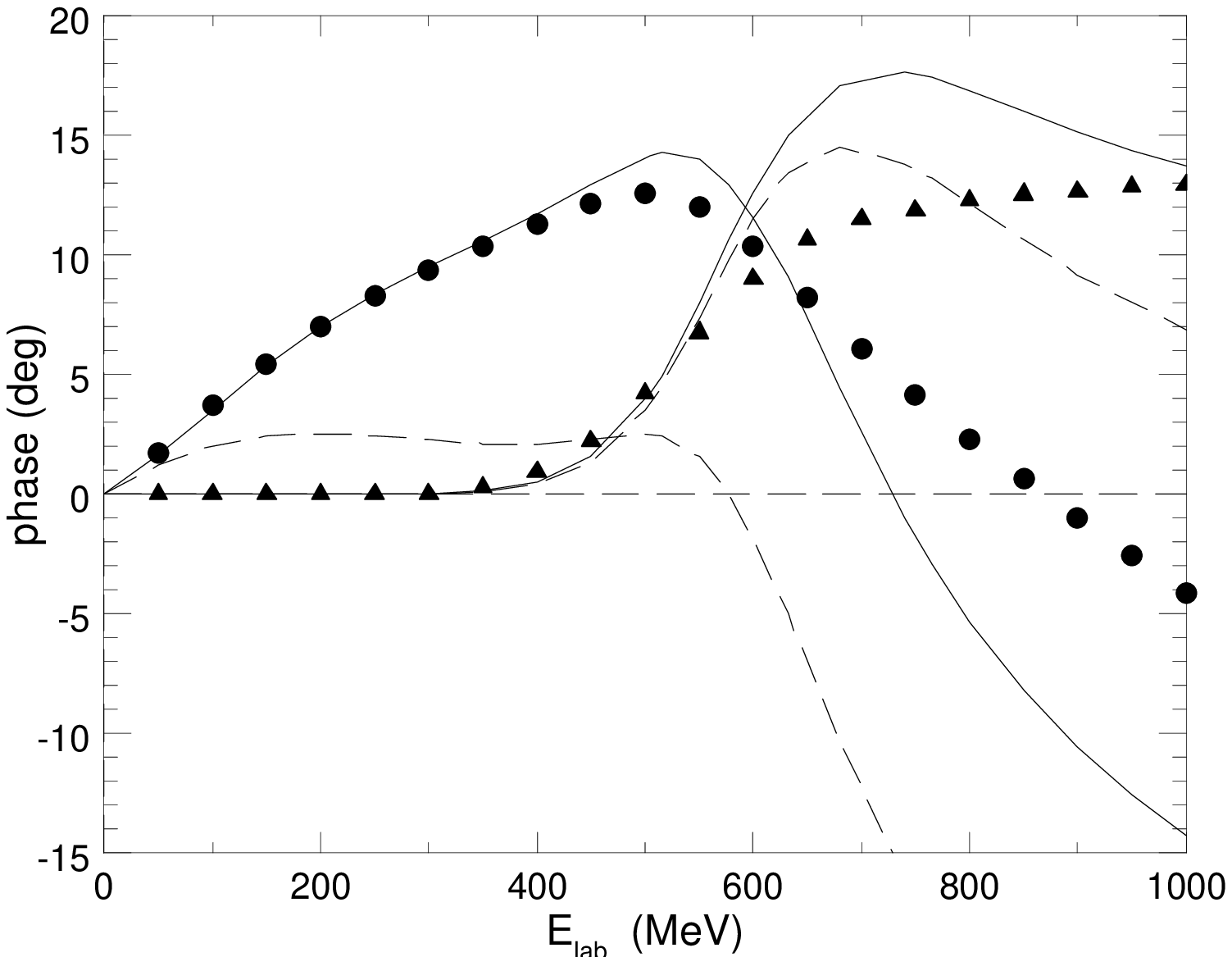}
\caption{The $^1D_2$ phase shift. The solid lines are the 
calculation  with $^5S_2 (N\Delta)$,
$^5D_2 (N\Delta)$ and $^5G_2 (N\Delta)$ admixtures and the filled 
circles and triangles are the experimental real and imaginary 
component of the phase shift extracted
from the energy-dependent fit to $pp$ data \cite{arndt}, 
respectively. The dashed curves only incorporate the 
$^5S_2(N\Delta)$ states as described in the text.
\label{dwave}
}
\end{figure}

First Fig. \ref{dwave} shows the behavior of the $^1D_2$ phase 
shift $\Re\delta$ together with its inelasticity  $\Im\delta$
including the possible  $^5S_2 (N\Delta)$,
$^5D_2 (N\Delta)$ and $^5G_2 (N\Delta)$ components in the
calculation (the solid curves). Here the data have been 
fitted to the real parts of phase shifts $\Re\delta$ 
\cite{arndt} below 
inelasticity (below 300 MeV), first starting with the 
phenomenological Reid potential  \cite{reid} and making 
the modifications to avoid the double counting of the 
additional attraction from 
the explicit $N\Delta$ inclusion as explained in 
the previous section.\footnote{One can
follow these steps for the $^3F_3$ wave in detail from Fig. 6 of
Ref. \cite{width}.} The data of Ref. \cite {arndt} have been very
stable below 1000 MeV, so they have been deemed precise enough
especially compared with any meaningful deviations from the
calculations.

The agreement with the elastic data \cite{arndt} is nearly 
perfect in the fitted region (as also for other
partial waves). Then the resonance-like energy dependence
will arise above 400 MeV due to the coupling to the 
isobar excitations, most 
pronouncedly to $^5S_2(N\Delta)$, not by any further fitting. 
It can be seen that, compared with
the energy-dependent analysis of $pp$ scattering \cite{arndt},
above 300 MeV the results even slightly exaggerate resonant 
behavior still after the smoothing effect of the width inclusion.
The structure follows clearly perturbative argumentation:
attraction below the threshold at about 600 MeV and then a 
change of the sign into repulsion. This behavior is faster 
above 600 MeV
than in Refs. \cite{faassen,elsternew} perhaps due to the
kinematically suppressed $N\Delta$ width and, because, to leave
space for possible ``dibaryons'', here no optimization is 
done above pion threshold.

The inelasticities of 
Elster {\it et al.} \cite{elsternew}
are smaller than here, because they use the weaker quark model
$\pi N\Delta$ coupling in the $NN \rightarrow N\Delta$ 
transition. As already emphasized in Sec. \ref{single},
this transition strength is essential and is fixed 
externally from $pp\rightarrow d\pi^+$ in this paper. 
The quark model coupling gives also 20\% underestimate 
for this reaction even if it is used only for the transition 
potential (as it would appear in \cite{elsternew},
more if it is also used in the final production vertex
\cite{wilhelm}.

In general, the $S$-wave $N\Delta$ component is the most important
in reactions in this energy regime and it is favoured also by the decrease of the angular momentum \cite{dibarmass} in the $D$- to
$S$-wave transition as noted in Sec. \ref{single}. 
Incidentally, at distances most relevant for pion exchange, here
this decrease is associated with an effective decrease in the 
centrifugal potential comparable to the threshold $\Delta M$,
making the $N\Delta$ excitation effectively rather degenerate 
with the initial $NN$ state in this region, and therefore 
strongly boosting this transition.

More specifically,
the $p$-wave pions associated with the initial $^1D_2(NN)$ 
state account for about 80\% of the $pp\rightarrow d\pi^+$ 
total cross section in the peak region around 580 MeV 
(laboratory energy) \cite{hoftiezer}, and this affluence 
arises mainly from the $^5S_2(N\Delta)$ admixture.
Therefore, it might be of interest to check separately
and explicitly the importance of different $N\Delta$
configurations for this reaction as well as for scattering.
In Fig. \ref{dwave} the dashed curves show the effect of
neglecting the coupling to the higher $^5D_2(N\Delta)$ 
and $^5G_2(N\Delta)$ channels in the equations (\ref{ccm}). 
The result is very interesting in that the
higher excitations (with even much higher effective thresholds
\cite{centrif}) are, however, still very important and 
essential in
providing attraction to $\Re\delta_2$. A similar but even 
more drastic influence of $D$-wave $N\Delta$'s will be seen
later
in the case of the $^1S_0(NN)$ partial wave. 

In contrast, the
effect of the higher $N\Delta$ states
to $\Im\delta_2$ is very small except fairly above the
$N\Delta$ threshold, reflecting the fact of the diminished width
due to the smaller free phase space for the $\Delta$ decay
\cite{width}.
Clearly the fast changing resonant behavior has moved to 
lower energies also with much lesser attraction.
 It should be noted that, to keep comparison
simpler in this calculation of the dashed curves, no further
readjustments for the $NN$ potential or the $N\Delta$ widths
were made. Also a possibly interesting finding for 
$pp\rightarrow d\pi^+$ was that the cross section with
$^5S_2(N\Delta)$ alone in this exercise
was in fact larger than the original, due to
destructive interference between $^5S_2(N\Delta)$ and
the smaller component $^5D_2(N\Delta)$, as may be inferred 
from e.g. Ref. \cite{ppdpi}.

The missing attraction at the highest energies might possibly 
be achieved also by inclusion of some higher nucleon resonance
($N(1520)$ or $N(1535)$) or double $\Delta$ excitation, beyond 
the scope of this paper.

\subsection{$^3F_3$ wave}

\begin{figure}[tb]
\includegraphics[width=\columnwidth]{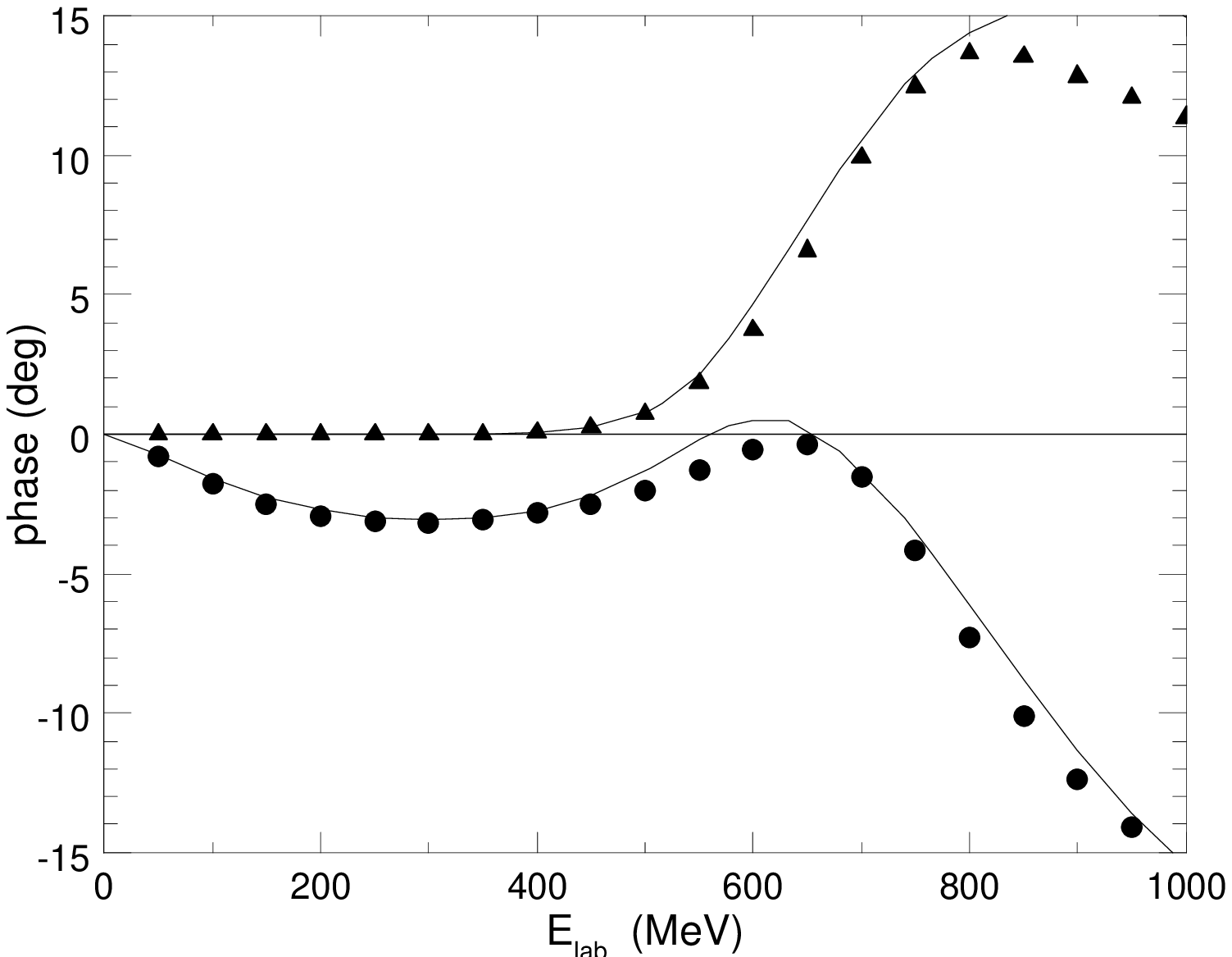}
\caption{As Fig. \ref{dwave} for the $^3F_3$ phase shift 
with $^5P_3 (N\Delta)$,
$^5F_3 (N\Delta)$ and $^3F_3 (N\Delta)$ admixtures. An earlier
calculation with only the $^5P_3 (N\Delta)$ component was presented
in \cite{width} emphasizing its role as a candidate dibaryon.
\label{fwave}
}
\end{figure}

Fig. \ref{fwave} shows the second important dibaryon candidate
$^3F_3$ starting from the $NN$ potential extension \cite{day}
of the Reid potential. 
Here the included $N\Delta$ configurations are 
$^5P_3 (N\Delta)$, $^5F_3 (N\Delta)$ and $^3F_3 (N\Delta)$ with
the first one being dominant due to the decrease of the orbital
angular momentum and consequently the smaller centrifugal 
potential in the $N\Delta$ system 
favouring it as argued already in Ref. \cite{dibarmass}.
This plot is very similar to Fig. 6 of Ref. \cite{width} where, 
however, only the ``dibaryonic'' state $^5P_3 (N\Delta)$ 
was included to emphasize its special influence and nature
as a ``dibaryon''.
The higher angular momentum states give minor attraction of
0.5--1.0 degrees above 500 MeV and, in comparison with 
Ref. \cite{width}, perhaps one might also argue a 
very slight broadening upwards in energy due to the stronger 
centrifugal effect in them. The ``knee'' at 600 MeV is better
reproduced here than in Refs. \cite{faassen,elsternew}
probably because of the dynamically calculated and reduced 
$N\Delta$ state
width and strong enough $NN\rightarrow N\Delta$ transition
potential.

This wave was found absolutely essential in 
successful production of differential observables in  
$pp\rightarrow d\pi^+$ \cite{ppdpi,improv} and it also accounts
for 10-20\% of its total cross section in the peak region. 
Probably the two dibaryonic states discussed so far dominate
also its isospin-cousin component in $pp\rightarrow np\pi^+$ 
at intermediate and high energies.
Furthermore, $^3F_3$ appears to give the largest individual
contribution at least to the total cross section of the
reaction $pp \rightarrow pp\pi^0$, with the final isotriplet 
$NN$ state, although overall this is significantly smaller than
the isosinglet.

\subsection{$^1 G_4$ wave}  \label{g4subsec}

\begin{figure}[tb]
\includegraphics[width=\columnwidth]{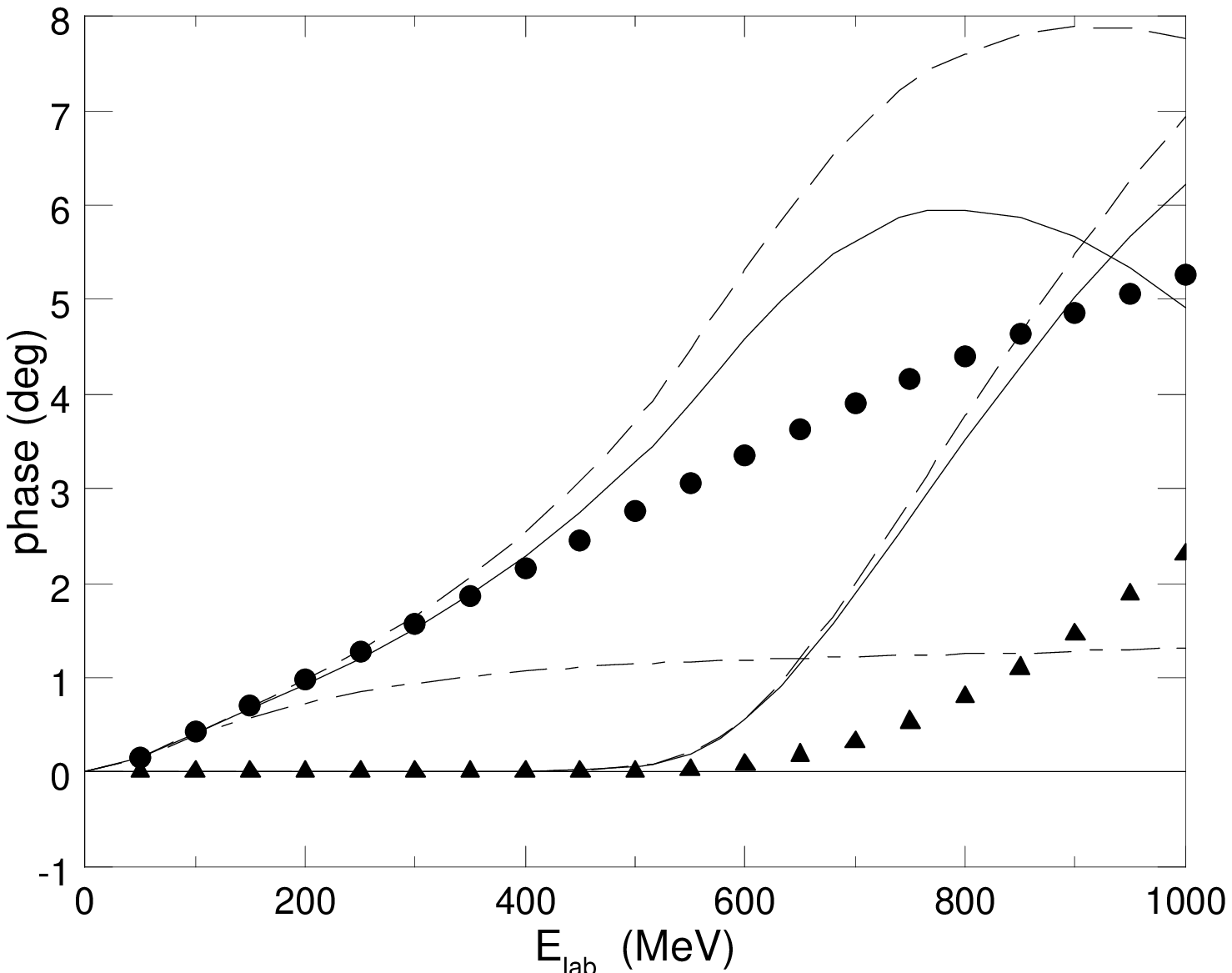}
\caption{Solid curves and data as in Fig. \ref{dwave} for the 
$^1G_4$ phase shift including $^5D_4 (N\Delta)$,
$^5G_4 (N\Delta)$ and $^5I_4 (N\Delta)$ admixtures.
Dashed: OPE used for the diagonal $NN$ potential but with $N\Delta$
admixture. Dashdot: only elastic OPE used overall (no $N\Delta$).
\label{gwave}
}
\end{figure}

The next potential $N\Delta$ dibaryon-like state, as argued in
Refs. \cite{dibarmass,centrif} and also experimentally suggested
in Yokosawa's review \cite{yokosawa}, $^1G_4(NN)$ is discussed 
in Fig. \ref{gwave} with the $^5D_4 (N\Delta)$,
$^5G_4 (N\Delta)$ and $^5I_4 (N\Delta)$ admixtures. Again,
computationally a clear broad resonance-like maximum is seen in the
phase shift at about 800 MeV laboratory energy, while in the data
the peak is missing. Also inelasticity is overestimated in
the calculated scattering.\footnote{For small inelasticity in the
presentation of this work as $\exp(-2\,\Im \delta)$ in the
$S$-matrix the
imaginary part of the phase $\Im \delta \propto \rho^2$ can
be more sensitive to variations than the $\rho$ parameter
of Ref. \cite{arndt}.}

Except for the $^1D_2$, in two-nucleon scattering the Argand 
diagrams of these states do not present any clear resonance 
behavior from phase shifts. Only the $^1D_2(NN)$ amplitude
crosses the imaginary axis slightly above 700 MeV (lab energy)
but with rather strong inelasticity \cite{width}. 
On the other hand, the $^3F_3$ amplitude 
remains on the left-hand side of the imaginary axis at all 
energies \cite{width}. 
The $^1G_4$ state has a calculated bump of $\Re\delta$ not seen
in the data. However, although the latter $NN$ amplitudes
may not show particular drastic features in actual pure
phase shift Argand diagrams, interestingly 
the $N\Delta$ {\em wave functions} do suggest resonant 
structures, both in magnitude and phase displayed 
for example in two Argand-like diagrams 
designed as probes for the  $^5D_4 (N\Delta)$ component in
Ref. \cite{centrif}, which may show up in reactions sensitive
to $N\Delta$ and involving overlaps with $N\Delta$ 
configurations.
Apparently the maximum in the phase shift here is due to 
constructive interference of the $N\Delta$ effect
with the attractive background from the $NN$ diagonal
potential (a modified $^1D_2$ Reid potential very 
similar to that used for Fig. \ref{dwave} was employed).

To study a little bit further this interplay 
in the $^1G_4$ wave, the dashed curves utilize instead of
the modified Reid potential as the diagonal $NN$ interaction 
the one pion exchange potential (OPE), expected to 
dominate peripheral waves. Although the fit (solid) is perfect,
also OPE as the diagonal potential is acceptable below 300-400
MeV (dashed). The overall phase $\Re\delta$
for this is slightly more attractive, increasingly
so for higher energies, and the maximum moves to higher energy.
This probably follows from the lack of the short range repulsion
present in the phenomenological potential. 

But more importantly,
it is very interesting and intriguing that the $N\Delta$
excitation is still very active and alive at such a high 
value of $L$, as seen in comparison of dashed curves vs. 
dash-dot (pure OPE), where the latter does not have 
this excitation at all.
This is probably due to the long (OPE) range of the transition
potential $V_{\rm tr}(r)$
itself and the favourable decrease of the centrifugal
repulsion in the change $^1G_4(NN) \rightarrow\ ^5D_4(N\Delta)$.

It might be argued that in high-$L$ peripheral waves only the
lightest meson $\pi$ could be exchanged and so only long range 
OPE could have any effect,  and excitation of
$\Delta$ would be of shorter range.  However, the transition
potential has the same range and once within that range the 
centrifugal $NN$ repulsion can be much larger than in the
$N\Delta$ system for $L' = L-2$. Therefore, the $N\Delta$
generation is not energetically an excitation at all, 
i.e. not like a heavier 
meson exchange. So once OPE is possible, so, to some extent
is $N\Delta$.

The OPE range is also essentially the range of the $N\Delta$ 
wave
function \cite{centrif}. Again, due to the difference in the
centrifugal potentials the $N\Delta$ admixture may not be
practically an excitation at all. The coupling 
effect looks even more drastic when comparing with the pure
OPE in the $NN$ wave without the $\Delta$-resonance generation
and the associated strong attraction
(dash-dot curve). Clearly, the $G$-wave is not peripheral enough
for the pure OPE to dominate alone at least above 300 MeV.
Quite evidently this raises the need to go further to even
more peripheral waves.
Below the inelastic threshold 300 MeV, the result here
(dashed vs. dash-dot) is numerically in perfect agreement with
the EFT calculation \cite{eft} in next-to-next-to-leading order
(NNLO) vs. leading order (LO).

The dipole form factor with the range parameter 
$\Lambda_\pi = 1000$ MeV has been included in the OPE's all 
the time.

\subsection{$J=0$ waves}

\begin{figure}[tb]
\includegraphics[width=\columnwidth]{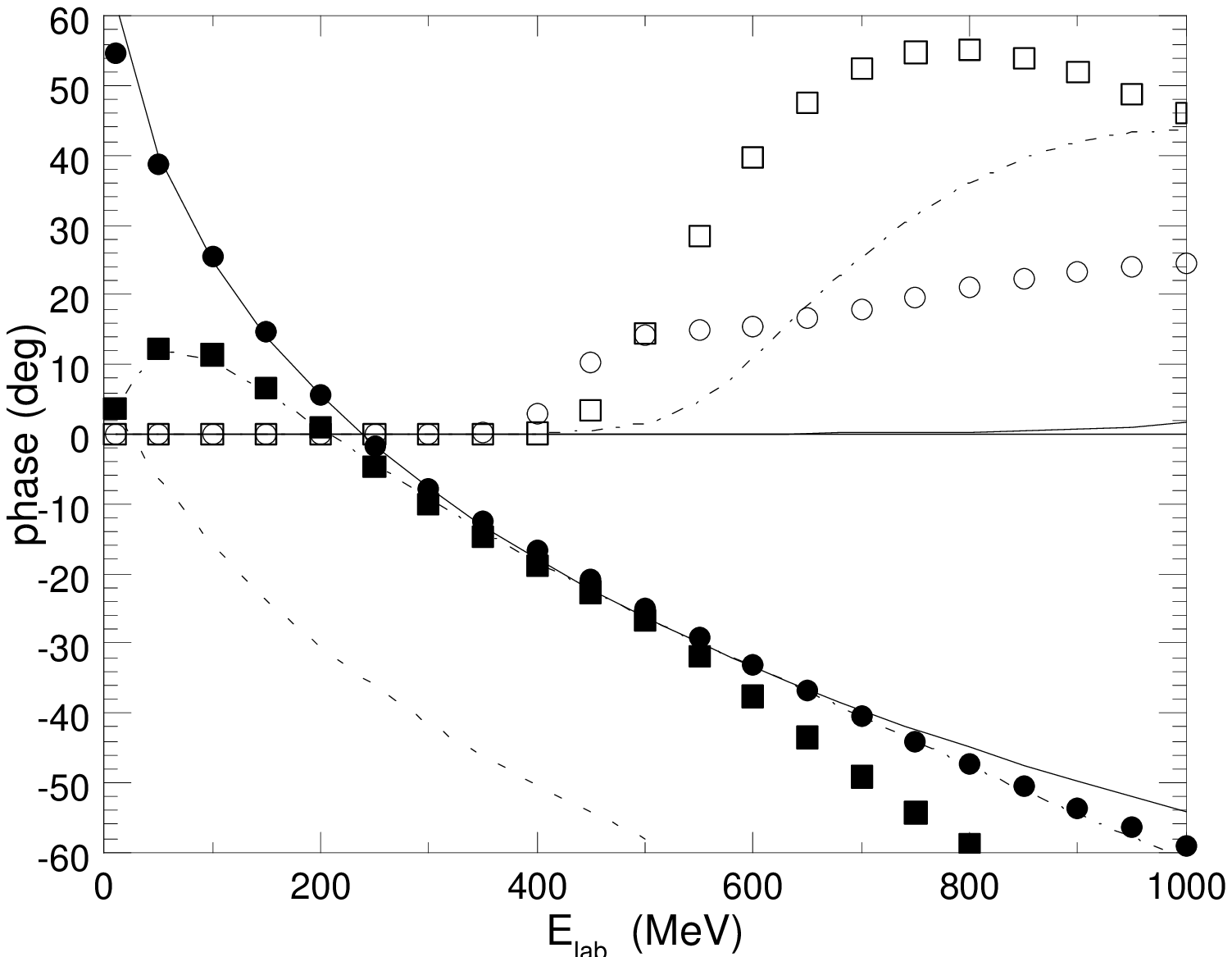}
\caption{$J=0$ phase shifts: solid curves $^1S_0$ and
dash-dot $^3P_0$. The circles show the analysis of Arndt
{\it et al.} \cite{arndt} for $^1S_0$ and the squares for $^3P_0$.
The filled symbols present the real parts and the hollow ones
the imaginary parts. For clarity, all imaginary parts 
$\Im \delta$ have been multiplied by 10. The dotted curve
shows the $^1S_0$ phase calculated {\it with} the 
double-counting correction but {\it without} the $N\Delta$
coupling.
\label{s0p0}
}
\end{figure}

After these somewhat peripheral waves it may be in order to
study similarly the opposite situation with the $J=0$ states, 
where the transition into dibaryonic
states with decreasing $L$ is not possible. Quite contrary,
the $^1S_0$ state is coupled only to the $D$-wave 
$^5D_0(N\Delta)$, which should involve some 300 MeV extra
centrifugal energy (plus some radial addition, too 
\cite{centrif}), bringing its effective excitation energy 
very high.

Fig. \ref{s0p0} presents the results for this wave by circles
and solid curves and for $^3P_0$ by squares and dash-dot curves.
A good agreement can be found for the real part in both
waves up to 800 MeV (well, up to 500-600 MeV for the $^3P_0$
wave).  In the $S$-wave the $N\Delta$ coupling 
is complemented with the repulsive double counting correction 
$770 \exp(-4\mu r) /(\mu r)$ MeV
to the Reid potential \cite{reid} - the phase has been fitted
between 50 and 300 MeV (the solid curve vs. filled circles).  
However, consistently with the arguments given earlier, 
the isobar excitation produces very little 
inelasticity - hardly distinguishable from the energy axis
even after multiplying by ten. This is due to the 
minuteness of the $^5D_0(N\Delta)$ width calculated with
the prescription of Refs. \cite{improv,width}. Likewise,
van Faassen and Tjon \cite{faassen} get also consistently
small inelasticity in this wave, whereas 
Elster {\it et al.} \cite{elsternew}
have this rising with energy. The latter may be, because in
an optical $NN$ potential it is difficult to include this 
state dependence on the $N\Delta$ effect. In fact, also the 
$^1S_0$ contribution to $pp\rightarrow d\pi^+$ cross section
is very small, in part due to destructive interference between
the $NN$ and $N\Delta$ components. The same seems to be true
for $pp\rightarrow pp\pi^0$ $s$-wave production (with $^3P_0$
final state nucleons).
To make the imaginary parts of the $NN$ phase shifts 
comparable in the scale, all of them have been
multiplied by 10. 

As a sideline, it is worth noting that, as a check, it was 
verified that in the low-energy limit the phase turns down at about 
1 MeV laboratory energy yielding for the singlet scattering length
a negative value 
$a_{\rm S} \approx -38$ fm. (Of course, in this context only the
sign matters.) Therefore, the possible presence of a bound
state and the ensuing extra node in the $NN$ wave function 
should not be the source for the smallness of the width.

As another sideline one might also note that, in spite of the
$^5D_0(N\Delta)$ state being unfavoured in reactions such as
$pp \rightarrow d\pi^+$ (due to destructive $NN$ and $N\Delta$
interference), its attractive effect may, however, be 
surprisingly strong. 
A quick calculation indicated that, with the direct $NN$ 
interaction totally discarded, the iteration of just 
the $N\Delta$ transition predicts a positive 
scattering length $a_{\rm S} \approx 2.3$ fm suggesting
that the $N\Delta$ component alone could support a bound state 
with a binding energy of  $E_{\rm B} \approx 8$ MeV.
To further illustrate the strength and importance of the 
$N\Delta$ configurations even in elastic scattering, the dotted
curve still shows the $^1S_0$ phase shift using the Reid
potential with the above
double-counting repulsion correction but {\it without}
including the $^5D_0(N\Delta)$ coupled channel. The gap
of 30--40 degrees between
the two results (solid vs. dotted) is massive and is one way of
viewing the significance of the $N\Delta$ coupling effect.

For the $^3P_0(NN)$ wave (coupled to $^3P_0(N\Delta)$), 
also with an attempted fit up to 300 MeV,
$\Re\delta$ agrees well with data up to about 600 MeV. Its
inelasticity agrees roughly with data 
in magnitude but would peak at higher 
energy. As noted before, this paper includes the inelasticity as 
the imaginary part $\Im \delta$ of the phase shift, which causes a 
different energy dependence from the $\rho$ parameter of \cite{arndt}
for small values. It may be of some interest to note that, 
especially for the $^3P_0$ state, also the large real part of 
the phase shift can have a strong effect in the inelasticity (with
the notation of Ref. \cite{arndt})
\begin{equation}
1 - \eta^2 = 
\frac{4 \tan^2\rho}{(1+\tan^2\rho)^2 + \tan^2\delta}\; ,
\end{equation}
bringing it down in spite of the seemingly very large values
of $\rho$.

The $J=0$ state pion reactions are somewhat superficially related
by their possibility to $s$-wave pion production by transitions
$^1S_0 \rightarrow\! ^3P_0 s$ and $^3P_0 \rightarrow\! ^1S_0 s$.
However, an explicit calculation suggests these reaction channels
to be significantly smaller than production in other $NN$ states
discussed here. Further, in their comparison, the former is nearly 
an order of magnitude smaller than the latter.

\subsection{$^3P_1$ wave}

\begin{figure}[tb]
\includegraphics[width=\columnwidth]{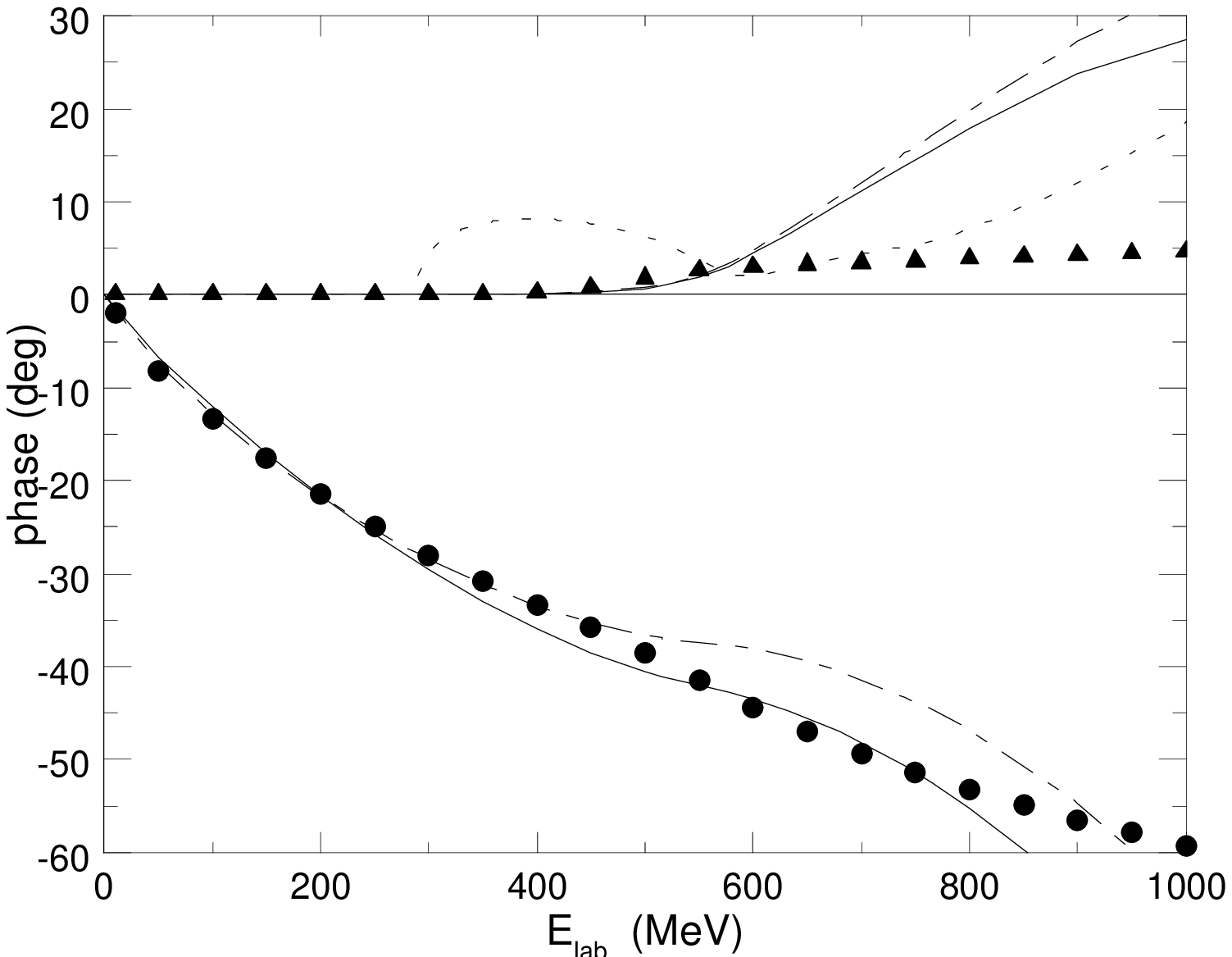}
\caption{$^3P_1$ phase shift: solid curves as described
in Ref. \cite{width} and earlier figures. 
The circles and triangles show the analysis of Arndt
{\it et al.} \cite{arndt}.
The dot-dash curve shows the modification described in 
the text, and the dotted one is 100 times  the sole 
contribution from $pp\rightarrow d\pi^+$ to inelasticity,
($s+d$)-wave pions.
\label{p1}
}
\end{figure}

Another $NN$ wave without a possibility of the favorable 
decrease of $L$ in the $N\Delta$ transition is $^3P_1$,
which has particular interest because of its significance 
in pion
production. Especially, due to $s$-wave pion rescattering 
in $s$-wave pion production, it dominates 
$pp\rightarrow d\pi^+$ and $pp\rightarrow np\pi^+$
at threshold.
Fig. \ref{p1} shows the results for this wave (solid
line) with a global fit to the phases of Ref. \cite{arndt} 
deep into pion production region. Generally there is a fair 
agreement of $\Re \delta$ with the smooth data behavior, 
quite enough for the originally intended 
calculation of pion production reactions
and isospin mixings. However, one may envision a tendency to
slight similarity with the $^3F_3$ wave (Fig. \ref{fwave}).
To emphasize this tendency one might try to improve the
already good fit by mimicking the curvature in the elastic 
region below 300 MeV (dash-dot line). Rather clearly now the
latter curve has as well a wide shoulder in the energy region 
600--800 MeV, similar to the case of $^3F_3$ also with 
$P$-wave $N\Delta$'s, but not really visible in the data for 
this wave.

The data indicate small and remarkably constant inelasticity 
$\Im\delta \approx 4^\circ - 5^\circ$
above 600 MeV, whereas the calculated results increase rather 
linearly with energy exaggerating it above 600 MeV. It is
probably useful to remember the behavior of $s$-wave pion
production in $pp\rightarrow d\pi^+$ \cite{ppdpi}.  
At threshold this 
$NN$ wave dominates, the cross section being linear in the 
pion c.m. momentum $q_\pi$, that is square root of the excess 
energy as is usual for the two-body final state phase space. 
However, this two-body reaction dominates the total
inelasticity of $^3P_1(NN)$ only up to 420 MeV.
Aut above this energy the scattering inelasticity 
calculated from the absorptive width quickly takes over 
so that already by 500 MeV two-body pion production
$pp\rightarrow d\pi^+$ is less than 10\% of the total. 
Moreover, due to destructive interference 
between direct production and pion $s$-wave rescattering
the cross section from this amplitude virtually vanishes 
around 600 MeV \cite{ppdpi}. 
Above that energy direct production 
from the axial charge part of the pion-baryon vertex wins
and the amplitude changes sign, but the $s$-wave pion cross
section remains still very small as seen already in 
Ref. \cite{ppdpi}. 
This wave is illustrated by the dotted curve in Fig. \ref{p1} 
(multiplied by the factor of 100, including also $d$-wave pions).
In the width prescription of Ref. \cite{width} the
phase space integral over intermediate $N\Delta$ 
momenta is augmented also by the  $pp\rightarrow d\pi^+$
cross section. 

To check the validity of this procedure
and to avoid problems with possible double counting
in explicit consideration of this reaction, the dash-dot 
curve now includes only the phase-space integrated width in the
calculation of the scattering inelasticity (and the elastic 
part of the phase shift). The $pp\rightarrow d\pi^+$ cross
section is then added explicitly by hand to yield the total
absorption cross section and finally to calculate the inelastic 
phase shift for the dash-dot curve from
\begin{equation}
1 - \eta^2 = \frac{k^2}{\pi (2J+1)}\; \sigma_{\rm abs}
  \label{inel}
\end{equation}
with $k$ the nucleon c.m. momentum.
In spite of the dominance of the $d\pi^+$ final state
at threshold (due to different phase space), its overall effect 
is very little for the present purpose, since actually also
the two-body reaction cross section is very small there. 
Above 600 MeV the inelasticity from this pion production
is less than 1\% of the scattering inelasticity so that 
its effect is very small whether it is included in the width
(imaginary potential) or explicitly added afterwards 
by Eq. (\ref{inel}) into the
total absorption cross section to calculate $\Im \delta$.
Namely the dotted curve shows 100 times its contribution.
It might still be noted that except for the minimum around
600 MeV this $^3P_1$-wave contribution 
to $pp\rightarrow d\pi^+$is still definitively 
dominated by $s$-wave pions over the also possible $d$-wave.

Is there a way to obtain the observed constancy of the 
inelasticity 
showing in the data? The structure of the $^3P_1$
contribution to  $pp\rightarrow d\pi^+$ (dotted curve) 
is intriguing as a total change from monotonously growing 
cross section at threshold (solid and dash-dot curves). 
In spite of this, the view of the 
$pp\rightarrow d\pi^+$ possibly changing
the trend of $\Im \delta$ towards a more constant value
seems hopeless due to the smallness of its $^3P_1$-wave
contribution. However, quite probably the importance of
$s$-wave pion rescattering at threshold also holds in
the isospin related three-particle final state in
$pp \rightarrow np\pi^+$. Furthermore, also the 
destructive competition between the axial charge
contributions could remain true producing a similar strong 
pion minimum for wave functions at any given final $NN$ energy. 
This might reduce
the smeared final state phase space integral of the cross
section to resemble better the constant inelastic data. 

True enough, also in the three-body $pp\rightarrow pp\pi^0$ reaction
there is a strong minimum in $s$-wave production 
around 550 MeV (however originating 
from the initial $^3P_0(NN)$ state), giving 
considerable constancy of the cross section up to near 600 MeV 
\cite{pppi0,meyer,impact} (though the cross section in this 
final state isospin arrangement is much smaller). The 
present phase  calculation, taking the width from integration
over intermediate $N\Delta$ momenta (and using as input for 
this the free $\Delta$ width) to simulate the three-body 
reaction \cite{width}, does not have these effects in. It 
does not require much imagination to consider how much the 
general outlook and the average of the dotted curve 
(Fig. \ref{p1}) would 
differ, if the minimum caused by the destructive 
interference would be smeared over in the phase space integral. 
However, it would require a totally independent
and different calculation of the three-body reaction.
Some hints in this direction may be found in 
Ref. \cite{budzanowski}.

\subsection{$^3P_2 -\! ^3F_2$ waves}

\begin{figure}[tb]
\includegraphics[width=\columnwidth]{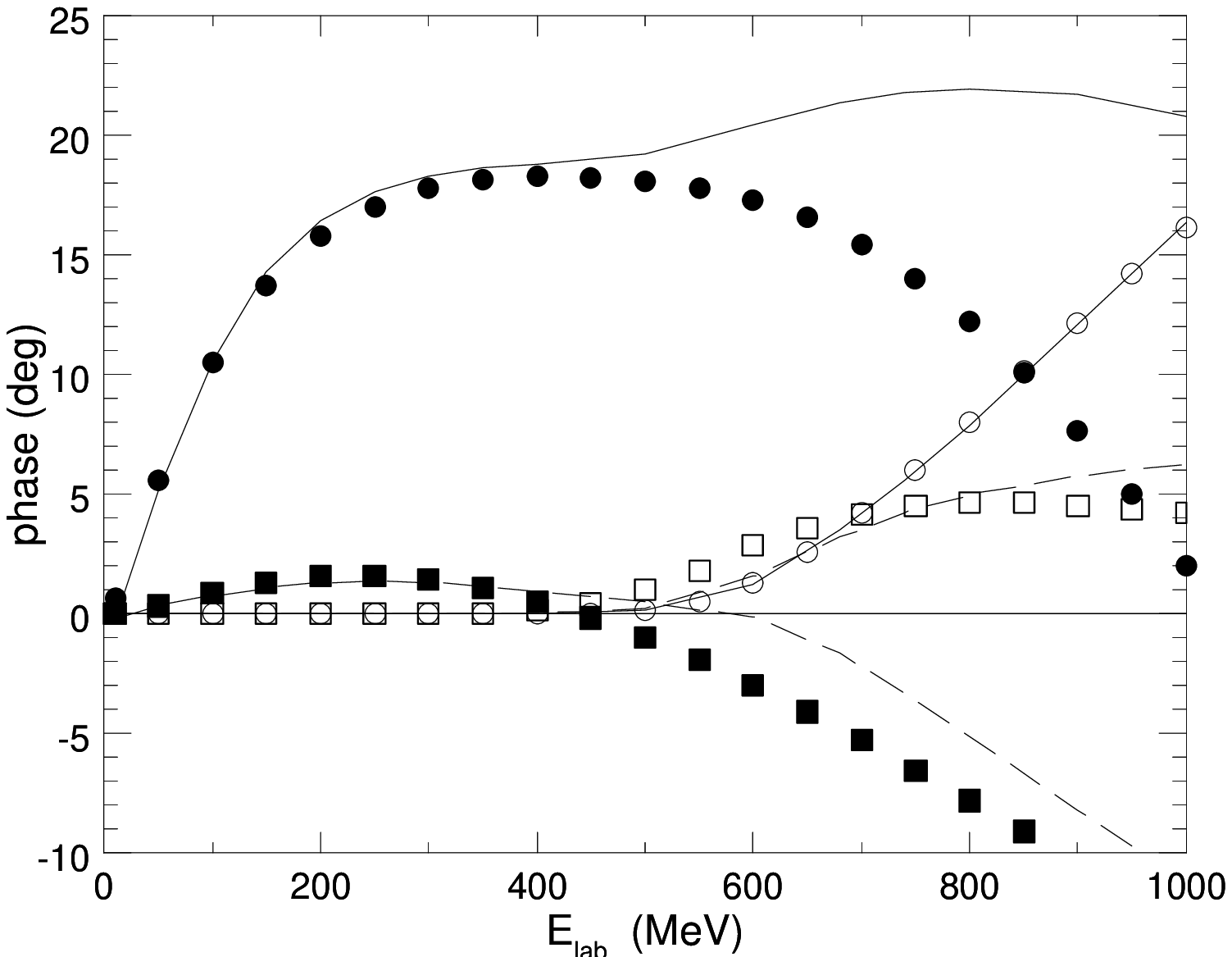}
\caption{$^3P_2 -\! ^3F_2$ phase shifts: solid curves the real 
and 
imaginary parts of the $P$-wave shift, the dashed curves for
the $F$ wave. Respectively,
the circles and squares show the analysis of Arndt
{\it et al.} \cite{arndt} transformed to the Stapp 
parametrization (\ref{stappmatrix}) (filled symbols 
for $\Re\delta$, hollow symbols for $\Im\delta$). 
The $F$-wave experimental inelasticity 
is multiplied by ten.
\label{pfphases}
}
\end{figure}

The tensor coupled case $^3P_2 -{^3F_2}$ is significantly more
involved \cite{arndt}. In the elastic region it is standard to
express the $2\times2$ $S$-matrix in the Stapp form \cite{stapp}
\begin{equation}
S = \begin{pmatrix}
\cos(2\epsilon)\exp(2i\delta_-) & 
  i\sin(2\epsilon)\exp[i(\delta_- +\delta_+)] \\
i\sin(2\epsilon)\exp[i(\delta_- +\delta_+)] &
\cos(2\epsilon)\exp(2i\delta_+)
\end{pmatrix} \; ,  \label{stappmatrix}
\end{equation}
where, in addition to the phase shifts $\delta_\pm$ in the
respective $\pm = J\pm  1$ $NN$ channels treatable 
as before in this paper, the mixing parameter $\epsilon$
opens the possibility of their mixing.
In this paper inelasticity has been implemented by complex
phase shifts, which automatically arise for complex potentials.
For convenience in analyses Ref. \cite{arndt} accommodates
inelasticity instead by adding a symmetric imaginary $2\times 2$ 
component into the otherwise real elastic $K$-matrix 
\begin{equation}
K = i (1-S)(1+S)^{-1} . \label{kmatrix}
\end{equation}
Thus, keeping this as the real part of the $K$-matrix, it is
generalized to the complex form
\begin{equation}
K =  \Re K + i\, \begin{pmatrix}
\tan^2\rho_-  &  \tan\rho_- \tan\rho_+ \cos\phi \\
 \tan\rho_- \tan\rho_+ \cos\phi  &  \tan^2 \rho_+  
\end{pmatrix} \; .   \label{inelk}
\end{equation}
However, I proceed in the present work by forcing inelasticity
into the Stapp form. Then, even the mixing parameter may become
complex, in contrast to the choice of \cite{arndt}.

In the phases of Fig. \ref{pfphases} the elastic region fit 
of $\Re\delta$
continues as essentially perfect somewhat above 400 MeV, but 
above 500 MeV deviates strongly especially for the $^3P_2$-wave
being far too attractive (solid curve vs. full circles). 
The same behavior at high energies occurs also in 
Refs. \cite{faassen,elsternew} (essentially constant 
$\Re \delta_- \approx 20^\circ$ from 200 MeV to 1000 MeV).
However, inelasticity in this
incident wave is excellent (the imaginary part of the phase, 
solid curve vs. hollow circles). 
The experimental phase parameters are extracted from 
the data of Ref. \cite{arndt} using their prescription for
transforming the $K$-matrix to the $S$-matrix elements.
In contrast, for the incident $^3F_2$-wave the elastic phase 
remains reasonable, but the imaginary part is tenfold as 
compared with the nearly elastic data. However, one should 
note the earlier warning about
the different sensitivity of the two representations for
small parameters ($\Im\delta$ vs. $\rho$) in this case.

\begin{figure}[tb]
\includegraphics[width=\columnwidth]{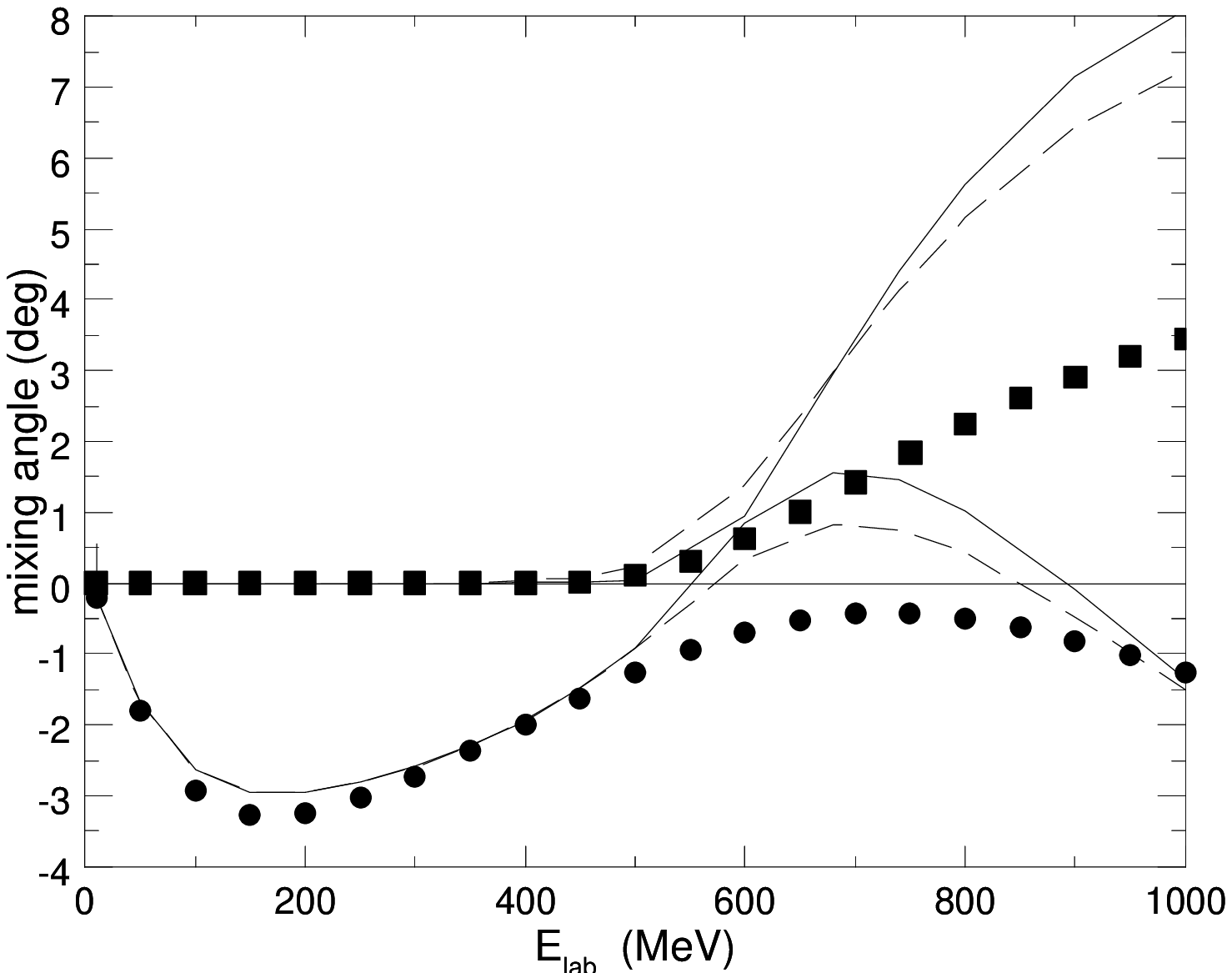}
\caption{The mixing angle $\epsilon$ as defined by 
Stapp {\it et al.} \cite{stapp} extended to inelastic region.
The circles ($\Re\epsilon$) and squares ($\Im\epsilon$) 
show the analysis of Arndt
{\it et al.} \cite{arndt}  forced into
the Stapp form. The solid curves show 
the results originating from the incident $P$-wave and
dashed curves from the $F$-wave.
\label{pfmixing}
}
\end{figure}

Fig. \ref{pfmixing} shows the $J=2$ mixing parameter $\epsilon$. 
In form $\Re \epsilon$ is very reasonable in comparison with
data (the lower curves vs. circles). However, the imaginary 
part (squares) may be surprising, not present in the original
definition (\ref{inelk}) of Ref. \cite{arndt}.
Nevertheless, from the inverse of Eq. (\ref{kmatrix})
\begin{equation}
S = (1 + i K)(1 - i K)^{-1}
\end{equation}
one can get the off-diagonal elements related by
(Eq. (7) of Ref. \cite{arndt})
\begin{equation}
S_0 = 2iK_0/D_K\; ,
\end{equation}
where $S_0$ (resp. $K_0$) is the off-diagonal element of the
$S$- and $K$-matrix and $D_K$ is the determinant of $(1-iK)$.
By definition (5) of Ref. \cite{arndt} above inelastic
threshold $K$ acquires both real and imaginary parts, and it is
perfectly plausible that the mixing parameter $\epsilon$ is
 complex. It is just
a consequence of the different definitions, and $\Im \epsilon$
may be rather directly (and simply) linked to $\rho$ at least for 
small mixing angles. In comparisons between theory and 
experiment, the {\it difference} between the upper curves and the 
squares is about the most relevant at this point in consideration
of how inelasticity is dynamically embedded in theory and shows up
in $\Im\epsilon$. It is probably worth noting that also van 
Faassen and Tjon get an imaginary component in $\epsilon_2$
remarkably the same size as here \cite{faassen}. 
The real part is somewhat 
better here, though qualitative trends are the same.

There is another perhaps strange point of interest and suspect. 
Numerically it was found that the exact values of $\epsilon$ 
depend on which of the two waves is the incident state (see 
solid vs. dashed curves in Fig. \ref{pfmixing}). This means that
the $S$-matrix is not perfectly symmetric as normally 
expected.\footnote{The $K$-matrix of Ref. \cite{arndt} defined
in Eq. (\ref{inelk}) is symmetric by construction.}
Apparently this feature has to do with the numerical finding
reported earlier in Ref. \cite{width} that, 
in addition to the rather natural dependence on the
relative orbital angular momentum between the $\Delta$ and
the nucleon themselves in the intermediate state, also the 
external incident $NN$ angular
momentum matters in the calculation and the value of the 
effective $N\Delta$ width. As noted already in Ref. \cite{width},
the treatment of inelasticity did not start from explicitly 
Hermitian first principles, but used existing inelasticity
(width) of the $\pi N$ resonance, the $\Delta$, as the starting
assumption. However, this may be a practicable procedure for
implementing intuitively apparently physical effects in the
system. Thus there would be somewhat 
different absorption for the two different initial waves. 
If this difference was taken off,
the $S$-matrix came out as symmetric (as in the formulation 
of Ref. \cite{arndt}). 
 Also, for the applicability of 
the theory it may be happy that this effect is actually small 
as compared even
with the (already rather small) total values of $\epsilon$.

\begin{figure}[tb]
\includegraphics[width=\columnwidth]{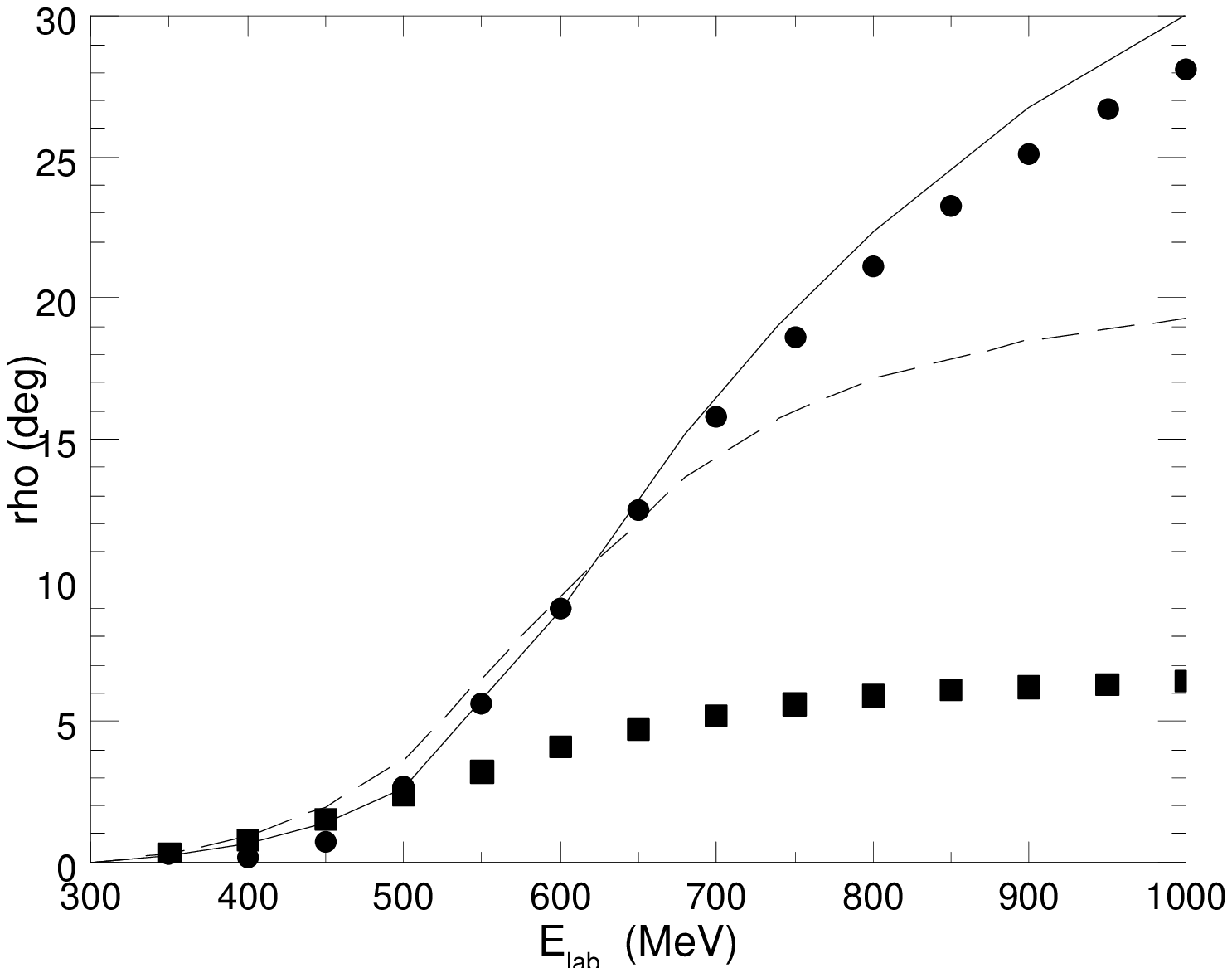}
\caption{The inelasticity parameter $\rho_\pm$ as defined 
by Arndt {\it et al.} \cite{arndt}. The $P$-wave solid vs. 
circles, the
$F$-wave dashed vs. squares. This may be compared with the
$\Im\delta$'s of Fig. \ref{pfphases} in the Stapp representation.
\label{rho}
}
\end{figure}

Finally, just for completeness and easier comparisons 
Fig. \ref{rho} presents the inelasticity parameter $\rho$
as defined by Arndt {\it et al.} \cite{arndt} easily 
obtainable by its definition (\ref{kmatrix})
from the calculated $S$-matrix as
\begin{equation}
\tan^2\rho_- = \Re(1 + S_+ -S_- -S_-S_+ + {S_0^-}^2) / D_S \; .
\end{equation}
Here, in an awkward notation $S_0^-$ is the off-diagonal 
$S$-matrix element originating from the initial $-$ 
channel (respectively $+ \leftrightarrow -$) and
$D_S$ is the determinant of $(1+S)$. The real phase
parameters $\Re\delta_\pm$ are practically the same in both
representations, Stapp or Arndt.

\subsection{More peripheral waves}  \label{peripheral}
\begin{figure}[tb]
\includegraphics[width=\columnwidth]{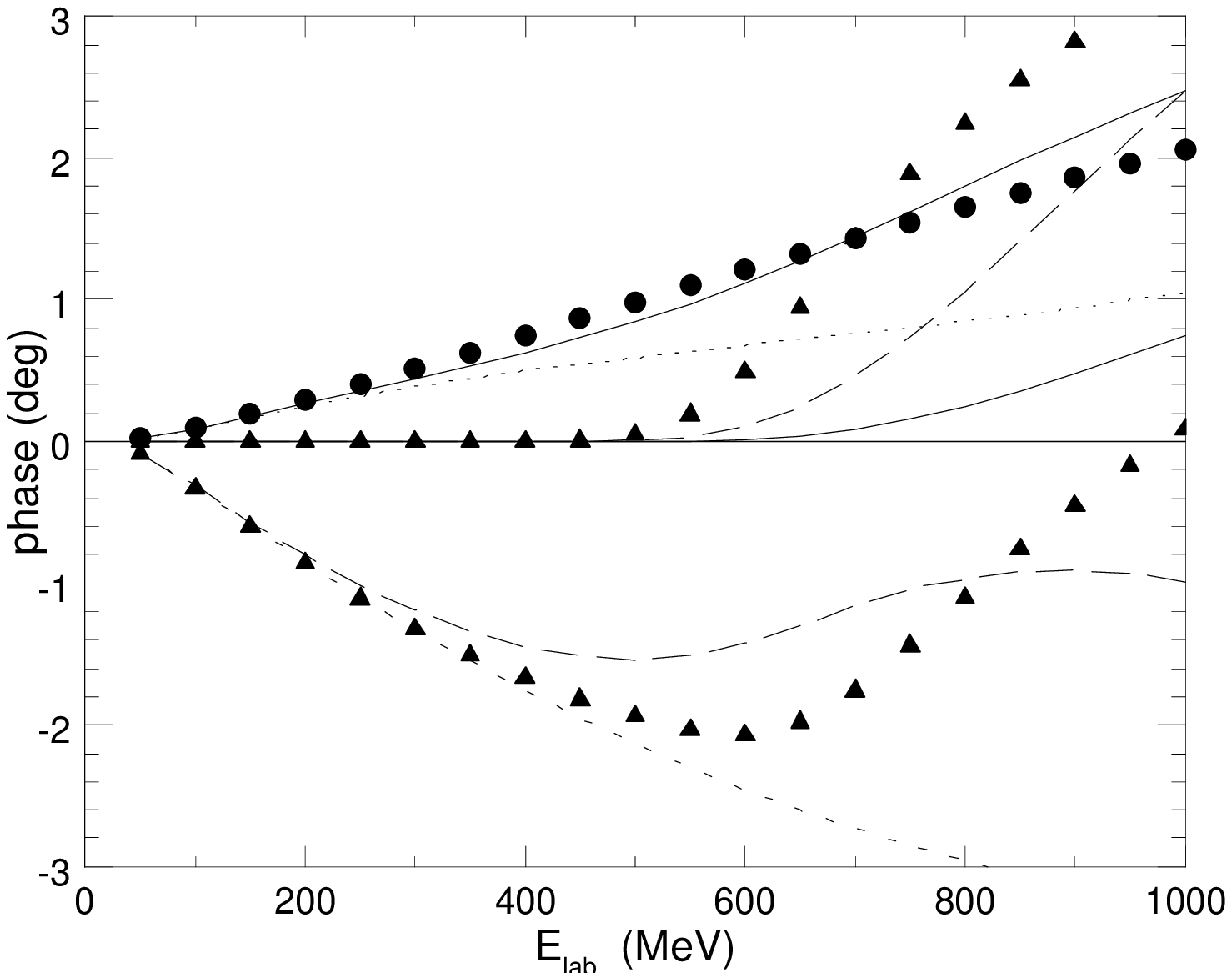}
\caption{Solid curves and circles show the $^1I_6$ partial wave
phase from the coupled channels calculation and from \cite{arndt}
(only $\Re\delta$ given for data). The dashed curves and triangles 
present the $^5H_5$ wave, and the dotted curves display the pure 
OPE result for both. \label{h5i6}
}
\end{figure}

In Subsec. \ref{g4subsec} the persistent  prominence of the 
attractive $N\Delta$ contribution even in the high $^1G_4$ 
partial wave is very suggestive that this
behavior may extend even higher. Fig. \ref{h5i6}
presents results for two more peripheral waves, making possible
a comparison of $^3H_5$ and $^1I_6$ waves including OPE together 
with $N\Delta$ or 
OPE alone. The dashed curves can be contrasted with the data of
Arndt {\it et al.} \cite{arndt} in the case of $^3H_5$. The model
gives quite similar behavior as seen for $^3F_3$, only at about 
300 MeV higher energy. The trend may also be seen in the data
(negative triangles), though the rise of the data at high 
energy is 
more pronounced and pointing towards even higher energies. The
predicted inelasticity is also tolerable. The solid curve 
vs. circles,
displaying the real part of the phase shift for the $^1I_6$ wave,
is perfectly satisfactory with coupled channels. Also the minor
imaginary part $\Im\delta$ is acceptable, though the data are
negligibly small and not shown.

One can make a very interesting comparison with the pure OPE
calculation (dotted curves). Clearly at and above the nominal 
$\Delta$ threshold isobar excitation is still well comparable
with OPE in such high angular momentum states. The effect 
is especially
outstanding and manifest in the destructive interference with
the repulsive triplet OPE both in $^3F_3$ and $^3H_5$ (also 
with a hint in $^3P_1$), challenging the old wisdom of the
OPE dominance at high $L$.

Similarly to the $^1G_4$ wave, below 300 MeV the deviation from 
OPE to full coupled channels (solid vs. dense dots for $^1I_6$
and dashed vs. sparse dots for $^3H_5$) is similar to the EFT
results of Ref. \cite{eft} from LO to NNLO.

\section{Conclusion}   \label{conclu}
The aim of the present paper has been to study $NN+N\Delta$
dynamics in the framework of coupled channels to produce 
resonant structures in $NN$ scattering without explicit 
input of energy dependencies, for reference and benchmarks 
to confront and compare with explicitly energy dependent 
fits, such as those with hypothesized dibaryons. 
Much of the necessary physics can be incorporated by
a formulation of coupled channels in terms of  
excitation of intermediate $N\Delta$ components. As already
seen at several points, 
this is not a new idea \cite{sugawara} and, in principle,
it has strong foundations in pion exchange: the pion coupling
to nucleons and $\Delta$'s is well established and should be
reconciled parallel with $NN$ scattering dynamics. Only after
considering this is it meaningful to involve other dibaryons.

The model, shortly outlined in Sec. \ref{single}. is 
substantiated by references to older work, in particular on
pion production, which is a natural association and a relative
to the present work. The relation to ``dibaryon'' fits is first 
to obtain as good a fit to the data as possible {\it below}
inelasticity and then leave the rest to depend on 
$N\Delta$  dynamics above the pion threshold with no 
supplemental degrees of freedom.

The desired features in the $\Delta$ excitation energy region 
are successfully predicted for the $NN$ waves which have also the 
most important resonant amplitudes in pion production in
$pp \rightarrow d\pi^+$ ($^1D_2$ and $^3F_2$, Figs. \ref{dwave}
and \ref{fwave}). The results overall agree quite well with 
earlier work on $NN$ scattering with $N\Delta$'s
\cite{faassen,elsternew}. Some of the differences may be 
identified in the state and energy dependence of the effective
$N\Delta$ width and the $NN\rightarrow N\Delta$ transition 
strength. This strength seems to be too weak in 
Ref. \cite{elsternew}.
Also the restriction of the phase shift fit only to the elastic
region below 300 MeV may be reflected into the resonant region 
as some deviation from models using global optimization.
This is a calculated choice to allow space for possible extraneous
dibaryons.

The more peripheral wave $^1G_4$ is seen to obtain still a major 
contribution from the $N\Delta$ coupling, even overpowering OPE, 
normally expected to overwhelmingly dominate (Fig. \ref{gwave}). 
This may be understood by the interplay with centrifugal 
effects: effectively the intermediate\linebreak 
$^5D_4(N\Delta)$ state is not energetically higher
than the $NN$ centrifugal barrier within the OPE range and thus is
not particularly an excitation. The calculated phase shift 
$\Re\delta_4$ displays rather a clear and wide maximum not seen 
in the data. The inelasticity 
$\Im\delta_4$ is somewhat overestimated in comparison
with Ref. \cite{arndt}, but this may be exaggerated
for small phases by the different parametrization. The
importance of the $N\Delta$ coupling prompted further
study of the peripheral waves in Subsec. \ref{peripheral}.

In the $J=0$ states the real phases come out rather good and also 
$\Im\delta$ for the $^3P_0$ is reasonable though not perfect
(Fig. \ref{s0p0}). The $^1S_0$ inelasticity does not agree 
with data, being much smaller, essentially negligible.
Calculationally the reason is the very small calculated 
$N\Delta$ width and the large centrifugal barrier 
in the intermediate $^5D_0(N\Delta)$ state essentially 
doubling the effective $N\Delta$ threshold \cite{centrif} 
(both are interrelated with each other). 
Also the  phase shift analysis result \cite{arndt}
for $\Im\delta$ is exceedingly small for such a low partial
wave, and one might again remind on the sensitivity of the
different parametrizations for small phases.
The calculated $^1S_0$ contribution to $pp\rightarrow d\pi^+$
is also very small, practically negligible in its total cross
section, well in line with the present results. Theoretically
it is interesting that the attractive strength of the explicit
$N\Delta$ is huge, 30--40 degrees from Fig. \ref{s0p0}, or 
seen in another way alone by itself enough to bind the $NN$.

In the $^3P_1$ incident wave a slight shoulder is predicted in
$\Re\delta_1$, not seen in the smooth data 
(Fig. \ref{p1}). Also the present
calculation overestimates the small and constant $\Im\delta_1$. 
However, possible improvements are suggested in the text from 
explicit experience in pion production (e.g. destructively
interfering pion $s$-wave rescattering \cite{ppdpi} missing
in the present calculation). After all, the inelasticities are
predominantly pion production.

In the tensor-coupled system $^3P_2-\! ^3F_2$ the real phases are 
reasonable, except for the $P$-wave $\Re\delta$ above 
600 MeV, which is much too attractive (Fig. \ref{pfphases}).  
The real part of the mixing angle
$\Re\epsilon_2$ compares reasonably with the data (which only
do give the real part) (Fig. \ref{pfmixing}).
However, the dynamic coupled channels 
model calculation also predicts an imaginary part for this.
An interesting peculiarity (due to the different definitions) is
getting an imaginary part for the Stapp mixing angle
$\epsilon_1$ from the real Arndt parameter $\rho$
(therefore only about half of it being of dynamical
origin, the other half coming from the transformation between
the two representations).
Another interesting, perhaps troublesome, result 
was that, parallel to earlier findings
\cite{improv,width}, also the angular momentum of
the incident external $NN$ state influences 
the widths calculated in the $N\Delta$ intermediate states and the 
$S$-matrix becomes slightly asymmetric.

Furthermore, even in higher peripheral partial waves $^3H_5$
and $^1I_6$ the $N\Delta$ was seen to contribute to the 
elastic parts of the phase shifts comparably with OPE and to
inelasticity perhaps alone by construction. 
One might raise the question whether this behavior continues 
without end, thus challenging the idea of neglecting
anything else but OPE in peripheral waves. Below pion threshold
the results for these peripheral waves agree numerically
with those of Ref. \cite{eft} using EFT.

Obviously much of resonances or resonance-like structures can be
obtained by coupled channels involving $N\Delta $ excitations,
even slight indications arose in waves where they are not seen 
in data.
The method appears to be a very reasonable, even successful 
starting point for producing realistic
wave functions to use in further calculations. The main shortcoming
may be that some additional interactions in strongly interacting
medium are not automatically included which affect inelasticity, 
such as e.g. $s$-wave 
rescattering of pions in the $^3P_1$ wave. But this is just 
what the ``further calculations'' mean, with some references
pointed out in the text. However, most of the $p$-wave pion
effects come already in with the dominant $\Delta$. In spite of
the aforementioned shortcomings, apart possibly from excessive
attraction in the $^3P_2$ wave above 500 MeV, the remaining 
deviations from data do not show particularly systematic and
pressing need for extraneous dibaryons.

\begin{acknowledgments}
I thank M. E. Sainio for useful comments.
\end{acknowledgments}

\end{document}